\documentclass[aps,superscriptaddress,twocolumn]{revtex4-2} 
\usepackage{feynmp}
\usepackage{amssymb}
\usepackage{epsfig}
\usepackage{epstopdf}
\usepackage{graphicx}
\usepackage{subfigure}
\usepackage{mathrsfs}
\usepackage{hyperref}
\usepackage{cleveref}
\usepackage{cases}
\usepackage{bbm}
\usepackage{xcolor}
\usepackage{tabularx}
\usepackage{array}
\usepackage{multirow}
\usepackage{CJK}
\usepackage{lmodern}
\usepackage{textcomp}
\usepackage{booktabs}
\usepackage{dcolumn}
\usepackage{makecell}
\usepackage{appendix}
\usepackage{threeparttable}

\begin{document}

\begin{CJK*}{GBK}{song}


\title{Critical behavior around the
fixed points driven by fermion-fermion interactions and disorders in the nodal-line superconductors}


\date{\today}

\author{Wen-Hao Bian}
\affiliation{Department of Physics, Tianjin University, Tianjin 300072, P.R. China}
\affiliation{School of Physics, Nanjing University, Nanjing, Jiangsu 210093, P.R. China}
\author{Jing Wang}
\altaffiliation{Corresponding author: jing$\textunderscore$wang@tju.edu.cn}
\affiliation{Department of Physics, Tianjin University, Tianjin 300072, P.R. China}
\affiliation{Tianjin Key Laboratory of Low Dimensional Materials Physics and
Preparing Technology, Tianjin University, Tianjin 300072, P.R. China}

\begin{abstract}

We systematically investigate the intricate interplay between short-range fermion-fermion interactions and disorder
scatterings beneath the superconducting dome of noncentrosymmetric nodal-line superconductors. Employing the renormalization group that
unbiasedly treats all kinds of potential degrees of freedom, we establish energy-dependent coupled flows for all
associated interaction parameters. Decoding the low-energy information from these coupled evolutions leads to the emergence of
several intriguing behavior in the low-energy regime. At first, we identify eight distinct types of fixed points,
which are determined by the competition of all interaction parameters and dictate the low-energy properties. Next, we carefully examine
and unveil distinct fates of physical implications as approaching such fixed points.  The density of states of quasiparticles displays a
linear dependence on frequency around the first fixed point, while other fixed points present diverse frequency-dependent behavior.
Compressibility and specific heat exhibit unique trends around different fixed points, with the emergence of
non-Fermi-liquid behavior nearby the fifth fixed point. Furthermore, after evaluating the susceptibilities of the potential states,
we find that a certain phase transition below the critical temperature can be induced when the system approaches the fifth fixed point,
transitioning from the nodal-line superconducting state to another superconducting state. This research would enhance our
understanding of the unique behavior in the low-energy regime of nodal-line superconductors.
\end{abstract}


\maketitle



%
%
%
%
%
%
%
%

\section{Introduction}

The study of superconductivity has sparked extensive attention from both experimental and theoretical perspectives, which
has emerged as a central and highly discussed subject in contemporary condensed matter physics.
Conventional superconductors typically manifest an isotropic $s$-wave superconducting gap without any nodal points surrounding the Fermi surface~\cite{Tinkham1996Book,Anderson1997Book,Larkin2005Book}.
This signals that the fermionic excitations are forbidden in the low-energy regime. The behavior of such superconductors aligns seamlessly with the principles of the celebrated BCS theory~\cite{BCS1957PR}.
In sharp contrast, the high-temperature cuprate superconductors (HTSCs) commonly feature a $d$-wave superconducting gap,
allowing for four gapless nodal points on the Fermi surface~\cite{Lee2006RMP,Ding1996Nature,Loeser1996Science,
Valla1999Science,Yoshida2003PRL,Ronning2003PRL}. This unique characteristic allows the presence of gapless fermionic quasiparticles, even at the lowest-energy limit, enabling their participation in various physical processes and triggering a number of critical behavior~\cite{Lee2006RMP}. Beyond these nodal-point superconductors,
there exists a class of three-dimensional (3D) compounds~\cite{Bonalde2005PRL,Izawa2005PRL,Tateiwa2005JPSJ,
Reid2010PRB,Reid2010PRL,Song2011Science,Watashige2015PRX,Reid2016PRL} that
are equipped with nodal-line points. These materials share a nodal structure reminiscent of cuprate HTSCs but
go beyond the restriction of four nodal points, as illustrated in Fig.~\ref{Fig_fermion_surface}.
In terms of terminology, these materials are generally designated as the nodal-line superconductors, including
the over-doped pnictides superconductors  $\mathrm{Ba(Fe_{1-x}Co_x)_2As_2}$~\cite{Reid2010PRB,Reid2010PRL}, FeSe~\cite{Song2011Science,Watashige2015PRX} and $\mathrm{(Ba_{1-x}K_x)Fe_2As_2}$~\cite{Reid2016PRL},
 as well as the heavy-fermion superconductors $\mathrm{CePtSi_3}$~\cite{Bonalde2005PRL,Izawa2005PRL,Tateiwa2005JPSJ},
UCoGe~\cite{Slooten2009PRL,Gasparini2010JLTP}. These compounds exhibit nodal-line structures, presenting intriguing possibilities for further study of superconductivity.

In principle, a plethora of physical degrees of freedom collaboratively shapes and determine the
low-energy physics of a certain system. Particularly, the substantial roles of electronic correlations and their interplay with other physical degrees of freedom have garnered considerable attention in elucidating the low-energy behavior and potential phase transitions of fermionic materials~\cite{Fradkin2009PRL,Vafek2012PRB,
Vafek2014PRB,Herbut2014PRB,Herbut2014PRL,Herbut2015PRB,Herbut2016PRB,Roy-Slager2018PRX,Roy2018PRX,Wang2018-2019}.
Besides, the presence of disorder scatterings in real compounds can also induce a number of singular behavior in the low-energy regime~\cite{Novoselov2005Nature, Castro2009RMP, Hasan2010RMP, Altland2002PR,
Fradkin2010ARCMP, Das2011RMP, Kotov2012RMP,Wang2011PRB, Wang2013PRB,
Aleiner2006PRL,Foster2008PRB,Lee2017arXiv,Nandkishore2017PRB,Wang-Nandkishore2017PRB,Nandkishore2014PRB,
Roy2017PRB-96,Roy1812.05615,Roy2016SR,Nersesyan1995NPB,Qi2011RMP,Stauber2005PRB,Wang2017QBCP,Wang2018-2019,
Wang2020PRB,Wang2021NPB,Mandal2018AP}. Considering the nodal-line superconductors armed with the unique topologies of Fermi surfaces~\cite{Sigrist1991RMP,Matsuda2006JPCM,Sur2016NJP}, the gapless fermionic excitations
can always be excited from the nodal lines. Such gapless excitations not only interact with each other~\cite{Vojta2000PRL,Vojta2000IJMPB,
Vojta2000PRB,Sachdev2011book,Vojta2003RPP,Lohneysen2007RMP} but also intricately entangle with disorder scatterings~\cite{Kotov2012RMP,Wang2011PRB, Wang2013PRB,
Aleiner2006PRL,Foster2008PRB,Lee2017arXiv,Nandkishore2017PRB,Wang-Nandkishore2017PRB,Nandkishore2014PRB,
Roy2017PRB-96,Roy1812.05615,Roy2016SR,Nersesyan1995NPB,Qi2011RMP,Stauber2005PRB,Wang2017QBCP,Wang2018-2019,
Wang2020PRB,Wang2021NPB,Mandal2018AP}, giving rise to a series of unconventional yet intriguing behavior as energy scales decrease~\cite{Castro2009RMP,Hasan2010RMP,Qi2011RMP,Das2011RMP,
Kotov2012RMP,Altland2002PR,Lee2006RMP,Fradkin2010ARCMP,Sachdev2011book}. Generally, these interactions and entanglements are
closely associated with the construction of ground states and the fates of low-energy physical implications~\cite{Castro2009RMP,Hasan2010RMP,Qi2011RMP,Das2011RMP,
Kotov2012RMP,Altland2002PR,Lee2006RMP,Fradkin2010ARCMP,Sachdev2011book}.
Consequently, an unbiased consideration of both fermion-fermion interactions and disorder scatterings becomes
imperative to capture the critical physical phenomena in the low-energy regime.
The delicate balance between these influences are expected to govern the emergence of unconventional behavior
and provide valuable insights into the understandings of nodal-line superconductors.

Inspired by these considerations, our current research is dedicated to unraveling the intricate consequences of the interplay between short-range fermion-fermion interactions and disorder scatterings on the critical physical behavior in the low-energy regime beneath
the superconducting dome of nodal-line superconductors.
In order to equally treat all these physical ingredients, we adopt the powerful renormalization
group (RG) approach~\cite{Wilson1975RMP,Polchinski9210046,Shankar1994RMP}. This approach enables us to collect all one-loop corrections
stemmed from the potential interactions and establish the coupled RG equations involving all relevant
interaction parameters.  After performing the numerical analysis, a lot of interesting results are extracted from the RG equations.

At first, we obtain eight distinct types of fixed points after analyzing the RG equations
of interaction parameters due to the interplay between fermion-fermion interactions
and disorder scatterings. These fixed points dubbed FP-$i$ with $i=1-8$ in Sec.~\ref{Sec_fixed_points}
substantially dictate the low-energy properties of the nodal-line superconductors.

Subsequently, the critical behavior of physical implications around these fixed points are carefully
investigated. To begin with, we find that the density of states (DOS) of quasiparticles displays a linear dependence on
frequency ($\omega$) and vanishes at $\omega=0$ as approaching FP-1. In comparison, approaching other fixed points leads to an initial
increase followed by a decrease in DOS with tuning up the frequency. Particularly, the competition of all interaction parameters
results in a non-zero finite DOS at $\omega=0$ and the anisotropy of fermion velocities can quantitatively modifies the behavior of DOS as well.
In addition, we realize that the behavior of compressibility ($\kappa$) on the chemical potential ($\mu$) shares the similar tenencies
with its DOS counterpart. Specially, $\kappa(\mu)\propto\mu$ around FP-1 but initially climb up and then decreases with increasing $\mu$ nearby other FPs.
Moreover, we notice that specific heat ($C_V)$ of quasiparticles gradually increases with the increase of temperature ($T$) and
approximately presents $C_V(T)\propto T^{2+\delta}$ in the vicinity of FP-$i$ with $i\neq5$ and $\delta\ll1$. In sharp contrast, it is noteworthy that as the
system approaches FP-5, the specific heat manifestly deviates from $C_V(T)\propto T^{2+\delta}$ and exhibits an approximately linear dependence on temperature,
signaling the emergence of non-Fermi-liquid behavior caused by the interplay of fermion-fermion interactions and disorder scatterings.

At last, we discuss the potential phase transitions in the low-energy regime as the system is tuned to
all distinct kinds of fixed points. After calculating and comparing the energy-dependent susceptibilities of potential candidates, we
find that the system can undergo a certain phase transition only when the system tends towards FP-5.
Such a phase transition renders the nodal-line superconducting state to another
superconducting state below the critical temperature under the interplay of
all interaction parameters. In particular, the competition between two momentum orientations of two
quaisparticles involved in the fermion-fermion interactions
plays an important role in pinning down the specific state induced by the phase transition.

The rest of this paper is organized as follows.
We introduce the microscopic model and establish the effective field
theory in Sec.~\ref{Sec_model}. Then, the RG equations of all fermion-fermion
interaction parameters and disorder strengths are presented in Sec.~\ref{Sec_RG_analysis},
from which the eight distinct kinds of fixed points are extracted. Next, as approaching
these fixed points, the critical behavior of physical implications and the
potential phase transitions due to the interplay of fermionic interactions and disorder scatterings are
presented in Sec.~\ref{Sec_critical_behavior} and Sec.~\ref{Sec_phase_transitions}, respectively.
Finally, we provide a brief summary in Sec.~\ref{Sec_summary}.

\section{Effective theory}\label{Sec_model}

\begin{figure}[htpb]
  \centering
  \includegraphics[width=2.5in]{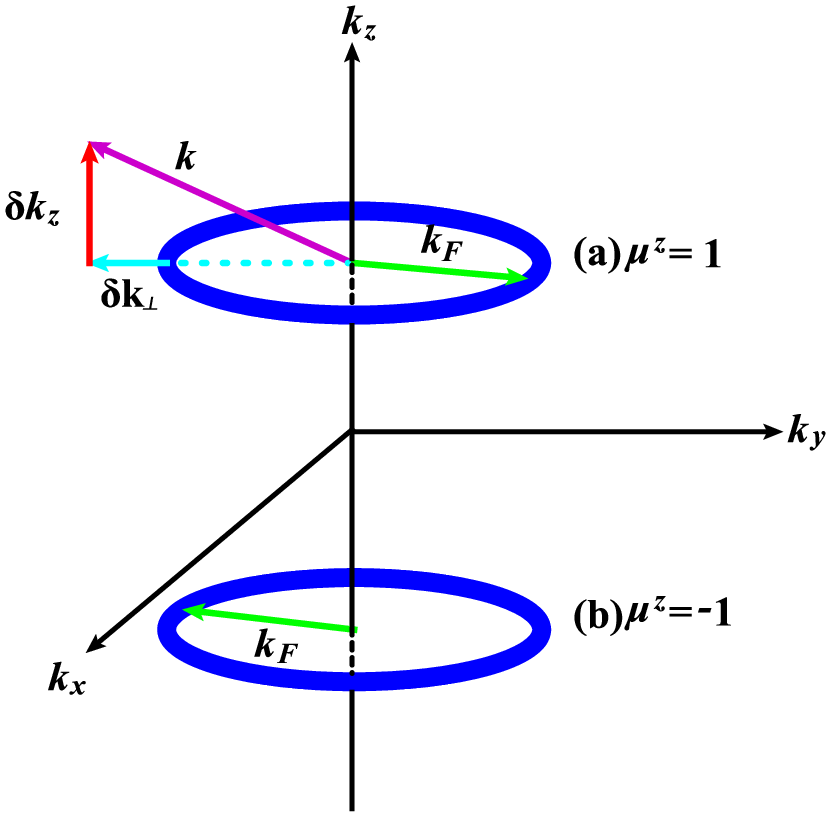}\\
  \caption{(Color online) Schematic illustration for the low-energy excitations around the nodal lines~\cite{Moon2017PRB}.} \label{Fig_fermion_surface}
\end{figure}

Within this work, we study the topological nodal-line superconductors with satisfying the
a common symmetry group $\mathcal{G}=C_{4v}\times\mathcal{T}
\times\mathcal{P}$ where $\mathcal{T}$ and $\mathcal{P}$ denote the time-reversal
and particle-hole symmetries, respectively~\cite{Matsuura2013NJP}.
Such materials possess two analogous nodal rings
dubbed by $\mu^z=1$ and $\mu^z=-1$ as schematically shown in Fig.~\ref{Fig_fermion_surface}~\cite{Moon2017PRB}.
Following the previous studies~\cite{Frigeri2004PRL,Brydon2011PRB,Moon2017PRB,BCW2023EPJP}, the effective non-interacting
Hamiltonian that characterizes the low-energy quasiparticle excitations can be written as
\begin{eqnarray}
H_{0}&=&\int\frac{d^3\mathbf{k}}{(2\pi)^3}\psi_{\mathbf{k}}^{\dag}(v_{z}\delta k_{z}
\Sigma_{03}+v_{p}\delta k_{\bot}\Sigma_{01})\psi_{\mathbf{k}}.\label{Eq_H_0-3}
\end{eqnarray}
Hereby, we employ the four-component spinor $\psi_{\mathbf{k}}^{\mathrm{T}}=
(c_{\mathbf{k},\uparrow},c_{-\mathbf{k},\uparrow},
c_{\mathbf{k},\downarrow},c_{-\mathbf{k},\downarrow})$ with the momentum $\mathbf{k}$
to describe the excited nodal fermions surrounding the upper ($\mu^{z}=1$) and lower ($\mu^{z}=-1$) nodal rings in
Fig.~\ref{Fig_fermion_surface}, and adopt $k_F$ to measure the radius of the nodal
line. In addition, the $4\times4$ matrix $\Sigma_{\mu\nu}$ is designated as
$\Sigma_{\mu\nu}\equiv\sigma_\mu\otimes\tau_\nu$ with $\mu,\nu=0,1,2,3$,
where the Pauli matrices $\tau^{1,2,3}$ and $\sigma^{1,2,3}$
apply to the particle-hole space and spin space, as well as $\tau^0$ and
$\sigma^0$ represent the identity matrix.
Moreover, $\delta k_{z}$ and $\delta k^2_\perp=\delta k^2_x+\delta k^2_y$ correspond to the transfer
momenta of the low-energy fermionic excitations in the $k_z$ direction and \emph{$k_{x}-k_{y}$} plane
around the nodal line, respectively~\cite{Moon2017PRB,BCW2023EPJP}. With this free Hamiltonian, we obtain the
noninteracting action as
\begin{equation}
S_0\!=\!\!\!\int\!\!\frac{d^3\mathbf{k}d\omega}
{(2\pi)^4}\psi^\dagger_{\mathbf{k},\omega}
(-i\omega+v_{z}\delta k_{z}\Sigma_{03}+v_{p}\delta k_{\bot}\Sigma_{01})\psi_{\mathbf{k},\omega},\label{Eq_S_0}
\end{equation}
which yields the free fermion propagator
\begin{eqnarray}
G_0(k)=\frac{1}{-i\omega+v_{z}\delta k_{z}\Sigma_{03}+v_{p}\delta k_{\bot}\Sigma_{01}}.
\end{eqnarray}
As only the physical degrees of freedom with small momenta
can be excited in the low-energy regime (\emph{$|\delta\mathbf{k}|\ll k_F$}), we hereafter employ the following
approximation suggested in Ref.~\cite{Moon2017PRB},
\begin{eqnarray}
\int\!\frac{d^3\mathbf{k}d\omega}{(2\pi)^4}\approx\int\frac{d\delta k_z}{2\pi}
\!\int k_F\frac{d\delta k_\perp}{2\pi}\!\int \frac{d\theta_\mathbf{k}}{2\pi}\!\int\frac{d\omega}{2\pi}.\label{Eq_G_0}
\end{eqnarray}

In addition to the free action~(\ref{Eq_S_0}),
we then introduce the fermion-fermion interactions and disorder scatterings.
For the former, there are 16 kinds of short-ranged fermion-fermion interactions between the low-energy excitations
from nodal lines~\cite{Vafek2012PRB,Vafek2014PRB,Wang2017QBCP,Moon2017PRB,
Roy-Sau2016PRB,Roy-Saram2016PRB,Roy-Sau2017PRL,Roy2021PRB},
\begin{eqnarray}
S_{\mathrm{ff}}\!&=&\!\!\!\!\sum^3_{\mu\nu=0}\!\!\!\lambda_{\mu\nu}\!\!\prod^4_{j=1}\!\!
\int\!\!\frac{d^3\mathbf{k}_jd\omega_j}{(2\pi)^4}
\psi^\dagger_{\mathbf{k}_1,\omega_1}\!
\Sigma_{\mu\nu}\psi_{\mathbf{k}_2,\omega_2}
\psi^\dagger_{\mathbf{k}_3,\omega_3}\!\Sigma_{\mu\nu}
\psi_{\mathbf{k}_4,\omega_4}\nonumber\\
\!\!\!\!\!\!&&\times\delta(\mathbf{k}_1+\mathbf{k}_2
+\mathbf{k}_3-\mathbf{k}_4)\delta(\omega_1+\omega_2
+\omega_3-\omega_4),
\end{eqnarray}
where the vertex matrices $\Sigma_{\mu\nu}\equiv\sigma_\mu\otimes\tau_\nu$ as aforementioned and $\lambda_{\mu\nu}$
are employed to measure the strengths of fermion-fermion couplings
with $\mu,\nu=0,1,2,3$. As shown in our previous work~\cite{BCW2023EPJP},
it is convenient to utilize the Fierz identity~\cite{Herbut2009PRB,
Herbut2016PRB,Roy-Saram2016PRB} to reconstruct $S_{\mathrm{ff}}$ with only six independent
couplings, namely
\begin{eqnarray}
S_{\mathrm{int}}\!\!&=&\!\!\!\sum^6_{i=1}\!\lambda_i\!\prod^4_{j=1}\!\!
\int\!\!\frac{d^3\mathbf{k}_jd\omega_j}{(2\pi)^4}
\psi^\dagger_{\mathbf{k}_1,\omega_1}
\mathcal{M}_i\psi_{\mathbf{k}_2,\omega_2}
\psi^\dagger_{\mathbf{k}_3,\omega_3}\mathcal{M}_i
\psi_{\mathbf{k}_4,\omega_4}\nonumber\\
\!\!\!\!\!\!&&\times\delta(\mathbf{k}_1+\mathbf{k}_2
+\mathbf{k}_3-\mathbf{k}_4)\delta(\omega_1+\omega_2
+\omega_3-\omega_4),\label{Eq_S_int}
\end{eqnarray}
where  $\mathcal{M}_i$ corresponds to $\Sigma_{01}$, $\Sigma_{03}$,
$\Sigma_{23}$, $\Sigma_{30}$, $\Sigma_{31}$, and $\Sigma_{32}$
as well as $\lambda_i$ represents the associated strengths with $i$
running from $1$ to $6$.

As to the latter, we put the focus on a quenched, Gaussian white-noise disorder, which obeys the following
restrictions~\cite{Nersesyan1995NPB,Stauber2005PRB,Wang2011PRB, Mirlin2008RMP,Coleman2015Book,Roy2018PRX},
\begin{eqnarray}
\langle \mathcal{D}(\mathbf{x})\rangle=0,\hspace{0.5cm}\langle \mathcal{D}(\mathbf{x})
\mathcal{D}(\mathbf{x}')\rangle
=\Delta\delta^{(3)}(\mathbf{x}-\mathbf{x}').\label{Eq_S_d-d}
\end{eqnarray}
Hereby, $\mathcal{D}$ and $\Delta$ denote the impurity field and the concentration of the
impurity, respectively.  Resorting to the replica method~\cite{Anderson1975JPE,Lee1985RMP,Lerner0307471,
Wang2015PLA,Roy2018PRX}, the fermion-disorder interplay can be described as~\cite{Roy-Saram2016PRB,Roy2018PRX}
\begin{eqnarray}
S_{\mathrm{dis}}
&=&\sum_i\Delta_i\int d\mathbf{x}d\tau d\tau'
\psi^{\dagger}_\alpha(\mathbf{x},\tau)\Gamma_i\psi_\alpha(\mathbf{x},\tau)\nonumber\\
&&\times\psi^{\dagger}_\beta(\mathbf{x},\tau')
\Gamma_i\psi_\beta(\mathbf{x},\tau'),\label{Eq_S_dis}
\end{eqnarray}
where $\Gamma_1=\sigma_0\otimes\tau_0$,
$\Gamma_2=\sigma_0\otimes\tau_2$, $\Gamma_3=\sigma_0\otimes\tau_1$, and
$\Gamma_{4j}=\sigma_j\otimes\tau_3$ with $j=1,2,3$ serve as the random
charge, random mass, random axial chemical potential,
and spin-orbit scatters, respectively~\cite{Roy-Saram2016PRB,Roy2018PRX}.
Besides, $\alpha,\beta$ label the replica indices, and $\Delta_i$ is adopted to
specify the corresponding strength of fermion-impurity
coupling.

At the current stage, combining the free action~(\ref{Eq_S_0}) and the fermion-fermion interactions~(\ref{Eq_S_int}) as well as
the disorder scatterings~(\ref{Eq_S_dis}) directly gives rise to the effective action,
\begin{eqnarray}
S_{\mathrm{eff}}
&=&S_0+S_{\mathrm{int}}+S_{\mathrm{dis}},\label{Eq_S_eff}
\end{eqnarray}
which is considered as our starting point. With the effective action in hand,
we are going to carry out the RG analysis for all the
interactions appearing in $S_{\mathrm{eff}}$ and determine the potential fixed points in the lowest-energy limit,
around which the critical phenomena can be expected.

\section{RG analysis and fixed points}\label{Sec_RG_analysis}

\subsection{RG flow equations}\label{Sec_RG_equations}

Following the spirit of RG approach~\cite{Wilson1975RMP,Polchinski9210046,Shankar1994RMP}, we subsequently
consider the free term of effective theory as an initial fixed point that is invariant under the
RG transformation, which are followed by the RG rescaling transformations of momenta, energy and fermionic
fields~\cite{Shankar1994RMP,Kim2008PRB,Huh2008PRB,
She2010PRB,She2015PRB,Wang2011PRB,BCW2023EPJP}
\begin{eqnarray}
\omega&\rightarrow&\omega e^{-l},\label{Eq_RG-scaling-1}\\
\delta k_z&\rightarrow&\delta k_ze^{-l},\\
\delta k_\perp&\rightarrow&\delta k_\perp e^{-l},\\
\psi_{\mathbf{k},\omega}&\rightarrow&\psi_{\mathbf{k},\omega}e^{\frac{1}{2}\int^l_0dl(4-\eta_f)},\label{Eq_RG-scaling-2}
\end{eqnarray}
where the anomalous fermion dimension $\eta_f$ is expressed as~\cite{BCW2023EPJP}
\begin{eqnarray}
\eta_f=\mathcal{C}_{2}
(\Delta_{1}+\Delta_{2}+\Delta_{3}
+\Delta_{41}+\Delta_{42}+\Delta_{43}),
\end{eqnarray}
and it collects the intimate contributions from one-loop corrections due to the
interplay between fermion-fermion interactions
and disorder scatterings. Subsequently, starting from the effective action~(\ref{Eq_S_eff}),
we performed all the one-loop corrections to all
the interaction parameters in previous work~\cite{BCW2023EPJP}. With the one-loop corrections,
we are able to implement the standard momentum-shell RG analysis~\cite{Shankar1994RMP,Kim2008PRB,Huh2008PRB,
She2010PRB,She2015PRB,Wang2011PRB} and obtain the coupled RG flow equations
of all coupling parameters, which can be formally written as
\begin{eqnarray}
\frac{dv_{z,p}}{dl} &=& \mathcal{F}(v_{v,p},\lambda_{i},\Delta_{j}),\label{Eq_v_RG}\\
\frac{d\lambda_i}{dl} &=& \mathcal{J}(\lambda_{i},\Delta_{j}),\\
\frac{d\Delta_j}{dl} &=& \mathcal{K}(\lambda_{i},\Delta_{j}), \label{Eq_Delta_RG}
\end{eqnarray}
where the variable $l$ denotes the running energy scale with $l=0$ describing
the initial state~\cite{Shankar1994RMP,Kim2008PRB,Huh2008PRB,
She2010PRB,She2015PRB,Wang2011PRB,Wang2013PRB,Wang2014PRD,Wang2015PRB,Wang2017PRB,Wang2017QBCP,
Vafek2012PRB,Vafek2014PRB,Wang2007.14981,Wang2022NPB}.
Hereby, we designate the coefficients $\mathcal{F}$, $\mathcal{J}$, and $\mathcal{K}$ to
serve as the functions of  the fermion
velocities $v_{z,p}$ and fermionic interactions $\lambda_i$ with $i=1-6$ as well
as disorder scattering strengths $\Delta_j$ with $j=1,2,3,41,42,43$, respectively.
All these coefficients are explicitly expressed in Appendix~\ref{Sec_appendix-one-loop-RGEqs} for details.

\begin{figure}[htpb]
\centering
\includegraphics[width=3.5in]{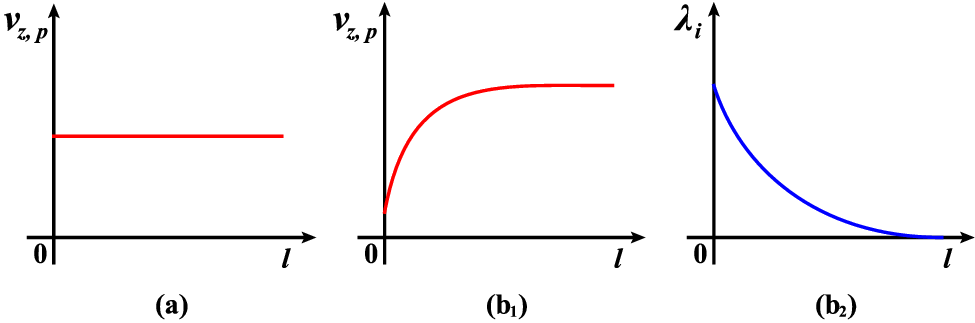}
\vspace{-0.5cm}
\caption{(Color online) Schematic tendencies diagrams of the fermion velocities
and fermion-fermion interactions as approaching (a) FP-1 for the noninteracting case and
(b) FP-2 for the clean limit case, respectively.}\label{class-FP-1-2}
\end{figure}

\begin{figure*}
\centering
\includegraphics[width=7.2in]{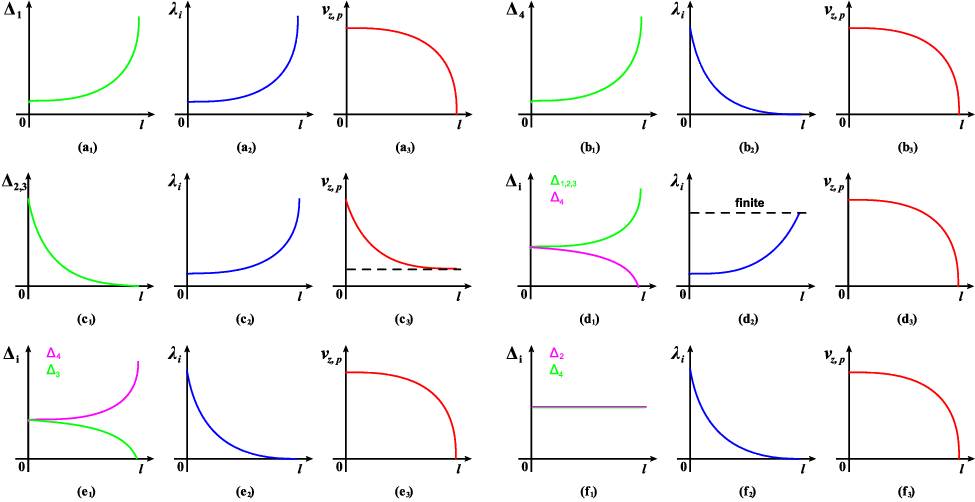}
\vspace{-0.5cm}
\caption{(Color online) Schematic tendencies of all interactions parameters as
approaching the other six types of fixed points: (a) FP-3 for the sole presence
of disorder $\Delta_1$, (b) FP-4 for the sole presence of disorder $\Delta_4$, (c) FP-5
for the sole presence of disorder $\Delta_2$ or $\Delta_3$, (d) FP-6
for the presence of all types disorders, (e) FP-7
for the presence of both disorder $\Delta_3$ and  $\Delta_{4}$,
and (f) FP-8 for the presence of both disorder $\Delta_2$ and
$\Delta_{4}$, respectively.}\label{class-FP-3-8}
\end{figure*}

Learning from these coupled RG equations~(\ref{Eq_v_RG})-(\ref{Eq_Delta_RG}), on one hand,  we find out that
the interaction parameters and fermion velocities may exhibit distinct fates in the low-energy regime
owning to the intimate entanglement among all the couplings.
On the other hand, the basic tendencies of the interaction parameters are principally can be dictated
by several fixed points in the parameter space (hereby, the FP is introduced to describe the features of
interaction parameters at a critical energy scale
characterized by $l_c$ beyond which the RG equations are invalid)
~\cite{Maiti2010PRB,Vojta2003RPP,
Roy2018PRX,Wang2017QBCP,Vafek2012PRB,Vafek2014PRB,
Wang2020PRB,Chubukov2012ARCMP,Chubukov2016PRX,Nandkishore2012NP,Wang2020NPB}.
As the system approaches such potential fixed points, the related physical obsverables are
generally expected to display unusual and critical behavior due to strong fluctuations around
the potential instabilities. Accordingly, it is of particular importance to determine all kinds of underlying fixed points
dictated by the RG equations~(\ref{Eq_v_RG})-(\ref{Eq_Delta_RG}), which we will
present in the next subsection~\ref{Sec_fixed_points}.

\subsection{Classification of fixed points}\label{Sec_fixed_points}

As aforementioned, the coupled RG flow equations~(\ref{Eq_v_RG})-(\ref{Eq_Delta_RG}) render all interaction parameters
appearing in Eq.~(\ref{Eq_S_eff}) mutually influenced as lowering the energy scale.  In order to capture their basic
tendencies, we carefully examine the behavior of RG equations with variation of the initial conditions of coupling parameters.
In principle, depending upon the initial conditions we figure out that the system would be attracted and hence governed by eight qualitatively different kinds of fixed points in the low-energy regime, which are schematically shown in Fig.~\ref{class-FP-1-2} and Fig.~\ref{class-FP-3-8}.

At first, there are two fixed points for the clean-limit case.
Specifically, under the non-interacting (free) situation, depicted in Fig.\ref{class-FP-1-2}(a),
fermion velocities remain unaffected as energy decreases, which is designated as the fixed point I (FP-1).
In contrast, while switching on the fermion-fermion interactions, we obtain the fixed point II (FP-2)
as illustrated in Fig.\ref{class-FP-1-2}(b), which shows that the fermionic
couplings gradually decrease at the low-energy limit,
and the fermion velocities $v_{p,z}$ progressively
climb up and become saturated at the lowest-energy limit.

In addition, for the sole presence of disorder, we find another three distinct fixed points depicted in
Fig.~\ref{class-FP-3-8}(a)-(c), which are nominated as the fixed point 3,4,5 (FP-3,4,5).
Learning from Fig.~\ref{class-FP-3-8}(a) and (b) for the presence of single disorder
$\Delta_1$ or $\Delta_4$, we notice that the disorder strength increases quickly but instead
the fermion velocities gradually decrease with eventually vanishing at a certain critical energy $l_c$.
Besides, the fermion-fermion couplings show a very distinct behavior for these two types of disorders,
they diverge at the $l_c$ with a single $\Delta_1$, while finally vanish
with a single $\Delta_4$. Besides, the single presence of disorder $\Delta_2$ or $\Delta_3$ share the similar behavior
as illustrated in Fig.~\ref{class-FP-3-8}(c). In such case, the
disorder strength progressively decrease as lowering the energy scale, and
the fermion velocities gradually decrease as well but ultimately converging to a finite non-zero value at a critical energy
scale. However, the fermionic couplings go up and tend to diverge.

Furthermore, there exist three fixed points named as FP-6,7,8 in Fig.~\ref{class-FP-3-8}(d)-(f)
for the simultaneous presence of multiple sorts of disorders.
As to this circumstance, we realize that the fermion velocities $v_{z,p}$ generally fall off as the energy scale is decreased
and go towards zero in the lowest-energy limit. In contrast, it is clear from the Fig.\ref{class-FP-3-8}(d)
that the fermionic interactions progressively increase and obtain a finite value as long as the disorder $\Delta_1$
is present. Otherwise, they flow towards zero without the participation of $\Delta_1$ shown in Fig.\ref{class-FP-3-8}(e)
and Fig.\ref{class-FP-3-8}(f). Considering the disorder strengths, $\Delta_{2,4}$ remain relatively
stable (Fig.~\ref{class-FP-3-8}(f)) but instead $\Delta_{3,4}$ diverges and tends to zero (Fig.~\ref{class-FP-3-8}(e)), respectively.
This therefore indicates that some of interaction parameters flow towards big values as approaching certain
fixed points. Generally, it is of interest to point out that the system usually exhibits non-trivial critical phenomena around such kinds
of fixed points~\cite{Vafek2012PRB,Vafek2014PRB,Roy2018PRX,Wang2017QBCP}.
In order to make the theory and our consideration under control within the perturbative theory, we are suggested to
rescale all interaction parameters by the dominant interaction parameter~\cite{Vafek2012PRB,Vafek2014PRB,Roy2018PRX,Wang2017QBCP}.
This gives rise to the rescaled FPs parameters, which consist of several small values. In other words, the coordinates of FP is equivalent
to $\{x_i(l_c)/\max\{x_i\}\}$ with the variable $x_i$ standing for all the interaction parameters $v_z,v_p,\lambda_{1-6},\Delta_{1,2,3,41,42,43}$
appearing in the effective action~(\ref{Eq_S_eff}).

In principle, these eight distinct kinds of fixed points encapsulate the basic low-energy physics of the effective theory~(\ref{Eq_S_eff}).
For the sake of completeness, the following studies will revolve around them and then uncover the distinct critical
properties caused by the interplay of fermionic interactions and disorders in the vicinity of all potential fixed points.

\begin{figure}[htpb]
  \centering
  \includegraphics[width=3.5in]{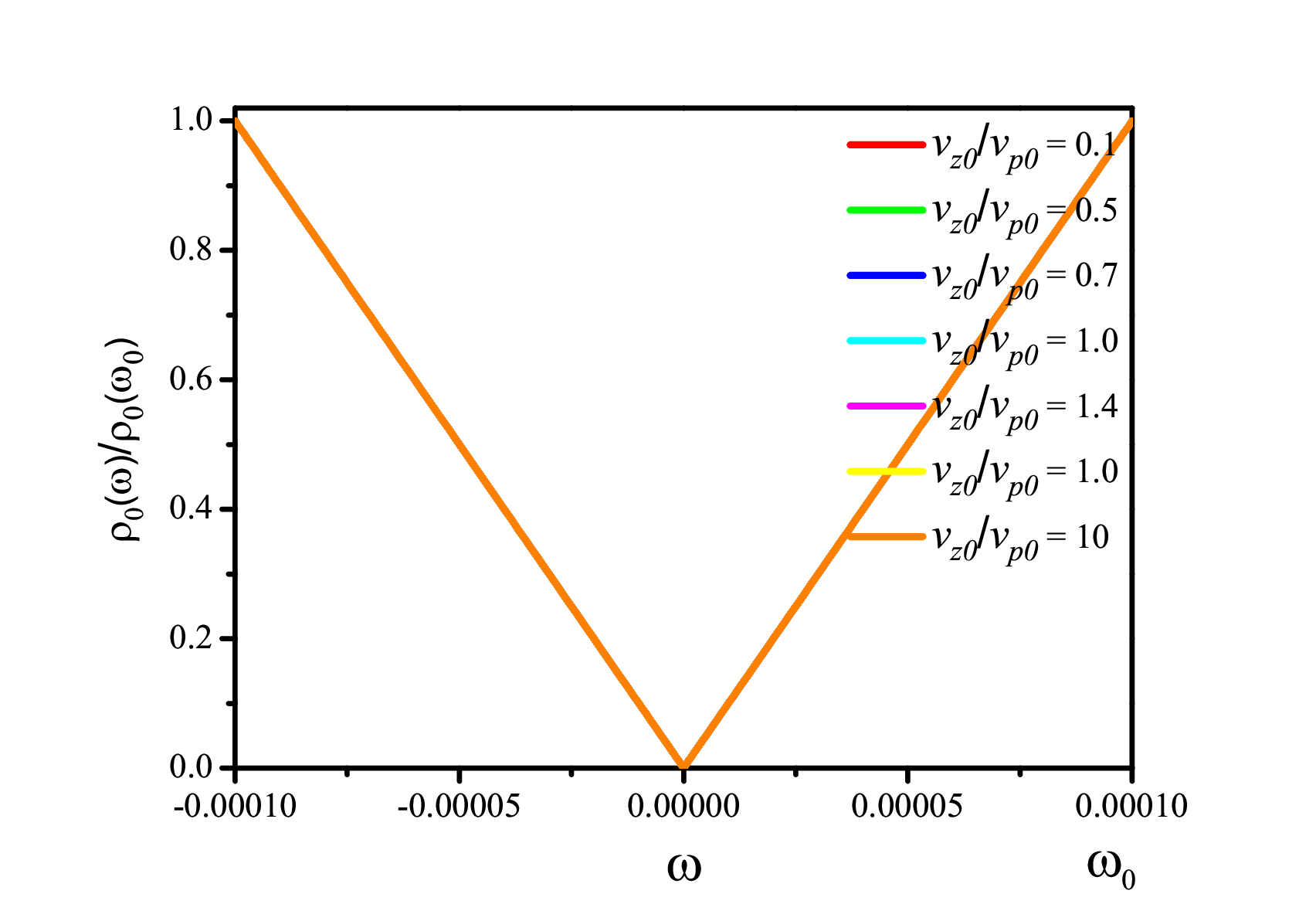}\\
  \vspace{-0.25cm}
  \caption{(Color online) Behavior of the DOS around the FP-1 (free case) for several
  representative values of fermion velocities ($v_{z0}/v_{p0}$) with $\omega_0=0.001$.}
  \label{DOS-FP-1}
\end{figure}

\section{Critical behavior as approaching fixed points}\label{Sec_critical_behavior}

In order to elucidate the critical effects resulting from the interacting terms in Eq.~(\ref{Eq_S_eff}),
we within this section will systematically investigate the
critical behavior of the observable physical quantities,
including density of states (DOS), compressibility, and specific heat of quasiparticles
when the system approaches the eight types of fixed points classified
in Sec.~\ref{Sec_fixed_points}. To this goal, we endeavor to derive the expressions of these physical
quantities. In principle, it is a very challenging task to derive the analytical expressions for
physical quantities directly from an interacting theory. Instead of delving into the exact
expressions, a practical approach is to extract the physical implications
from the renormalized fermionic propagator~\cite{Mahan1990Book},
which inherit the physical information decoded in the coupled the coupled RG equations
derived in Sec.~\ref{Sec_RG_equations}. Subsequently, we can examine their behavior as the system
flows towards the potential fixed points one by one.

\begin{figure}[htpb]
\centering
\subfigure[]{\includegraphics[width=3.5in]{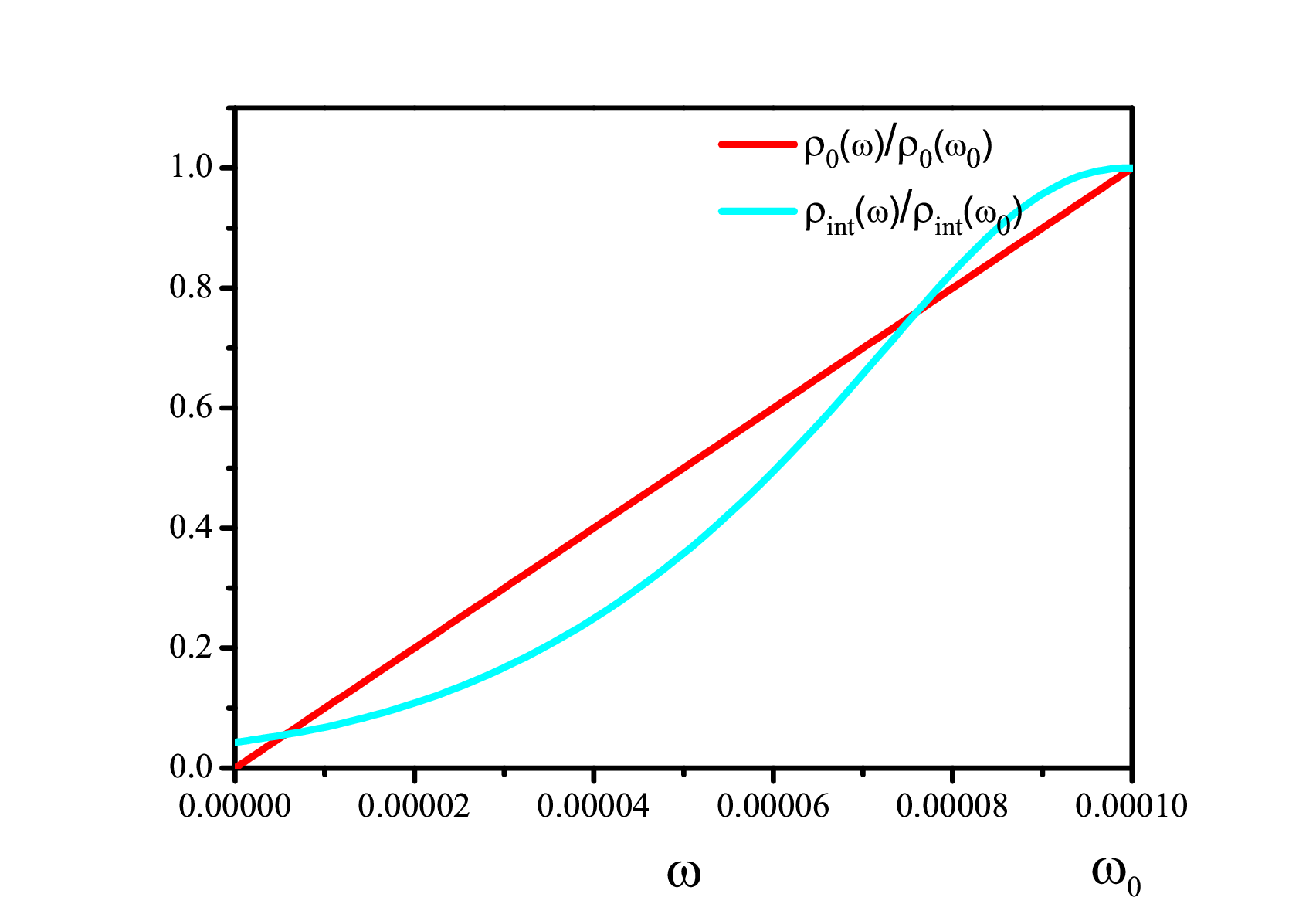}}\vspace{-0.50cm} \\
\subfigure[]{\includegraphics[width=3.5in]{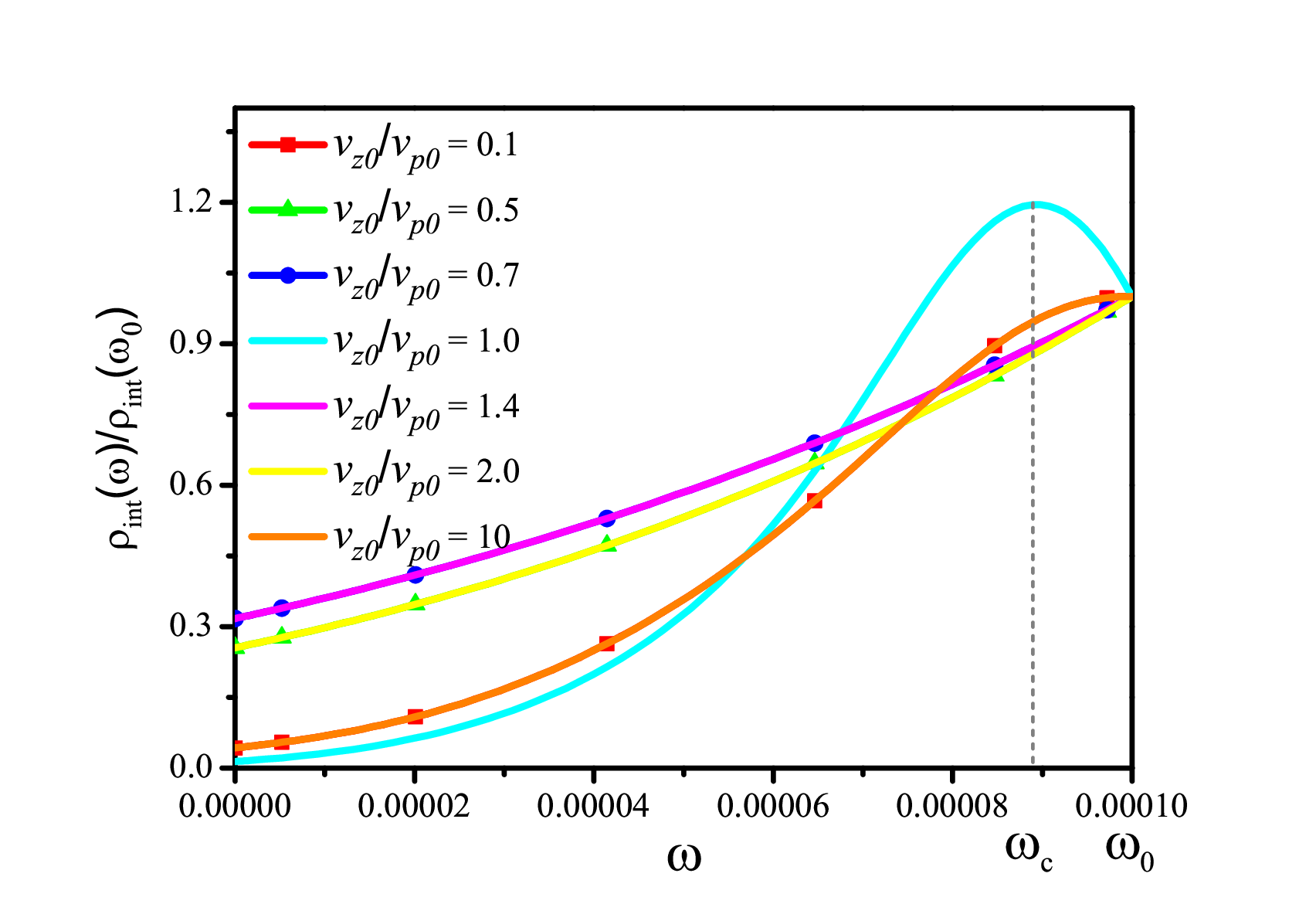}}
\vspace{-0.35cm}
\caption{(Color online) (a) Comparison of DOS around FP-1 and FP-2 at $v_{z0}/v_{p0}=0.1$,
and (b) Behavior of the DOS around  FP-2 (clean case)
for several representative values of fermion velocities ($v_{z0}/v_{p0}$) with $\omega_0=0.001$.}
\label{DOS-FP-2}
\end{figure}

\subsection{Density of states}\label{Subsection_DOS}

At the outset, we consider the density of states (DOS) of quasiparticles.
To this end, we begin with the noninteracting case in which the free propagator of quasiparticle is derived
in Eq.~(\ref{Eq_G_0}). After performing the analytical continuation, its retarded version
takes the form of~\cite{Mahan1990Book}
\begin{eqnarray}
&&G_0^{\mathrm{ret}}(\omega+i\delta,\mathbf{k})\nonumber\\
&=&
\frac{\omega+v_z k \sin\theta \Sigma_{03}+v_p k \cos\theta \Sigma_{01}}
{-\omega^2-i\mathrm{sgn}(\omega)\delta+v_z^2 k^2\sin^2\theta+v_p^2 k^2\cos^2\theta}.
\end{eqnarray}
It henceforth yields the related spectral function as follows~\cite{Kim-Kivelson2008PRB,Xu2008PRB,She2015PRB}.
\begin{eqnarray}
\mathcal{A}(\omega,\mathbf{k})
&=&
\frac{1}{\pi}\big\{\mathrm{Im}[G_0^{\mathrm{ret}}(\omega+i\delta,\mathbf{k})]\big\}\nonumber\\
&=&
\mathrm{sgn}(\omega)\left(\omega+v_z k\sin\theta\Sigma_{03}+v_p k \cos\theta\Sigma_{01}
\right)\nonumber\\
&&\times\delta\left(-\omega^2+v_z^2k^2\sin^2\theta+v_p^2k^2\cos^2\theta\right),
\end{eqnarray}
with which the DOS can be written as~\cite{Kim-Kivelson2008PRB,Xu2008PRB,She2015PRB}
\begin{eqnarray}
\frac{\rho_0(\omega)}{\Lambda_0 k_F}
&=&
N \int\frac{d^3\mathbf{k}}{(2\pi)^3}\mathrm{Tr}\left[\mathcal{A}(\omega,\mathbf{k})\right]\nonumber\\
&=&
\frac{N|\omega|}{2\pi^2}\alpha(v_z,v_p).\label{rho_0}
\end{eqnarray}
Here, $N$ serves as the fermion flavor, $\Lambda_0$ represents a cutoff momentum associated with
the lattice constant, and $k_F$ is the Fermi momentum.  We adopt the the transformation $\omega\to\omega/\Lambda_0$
to make it more compact. Besides, $\alpha(v_z, v_p) \equiv \int_0^{\pi}d\theta \big(v_z^2\sin^2\theta+v_p^2\cos^2\theta\big)^{-1}$
with $0 < |\omega|/\sqrt{v_z^2\sin^2\theta+v_p^2\cos^2\theta} < 1$ and $\omega\in(-1,1)$.
It is necessary to highlight that $v_z$ and $v_p$ are certain constants in the absence of
the interactions and the DOS apparently vanishes at the Fermi level.

\begin{figure}[htpb]
\centering
\includegraphics[width=3.5in]{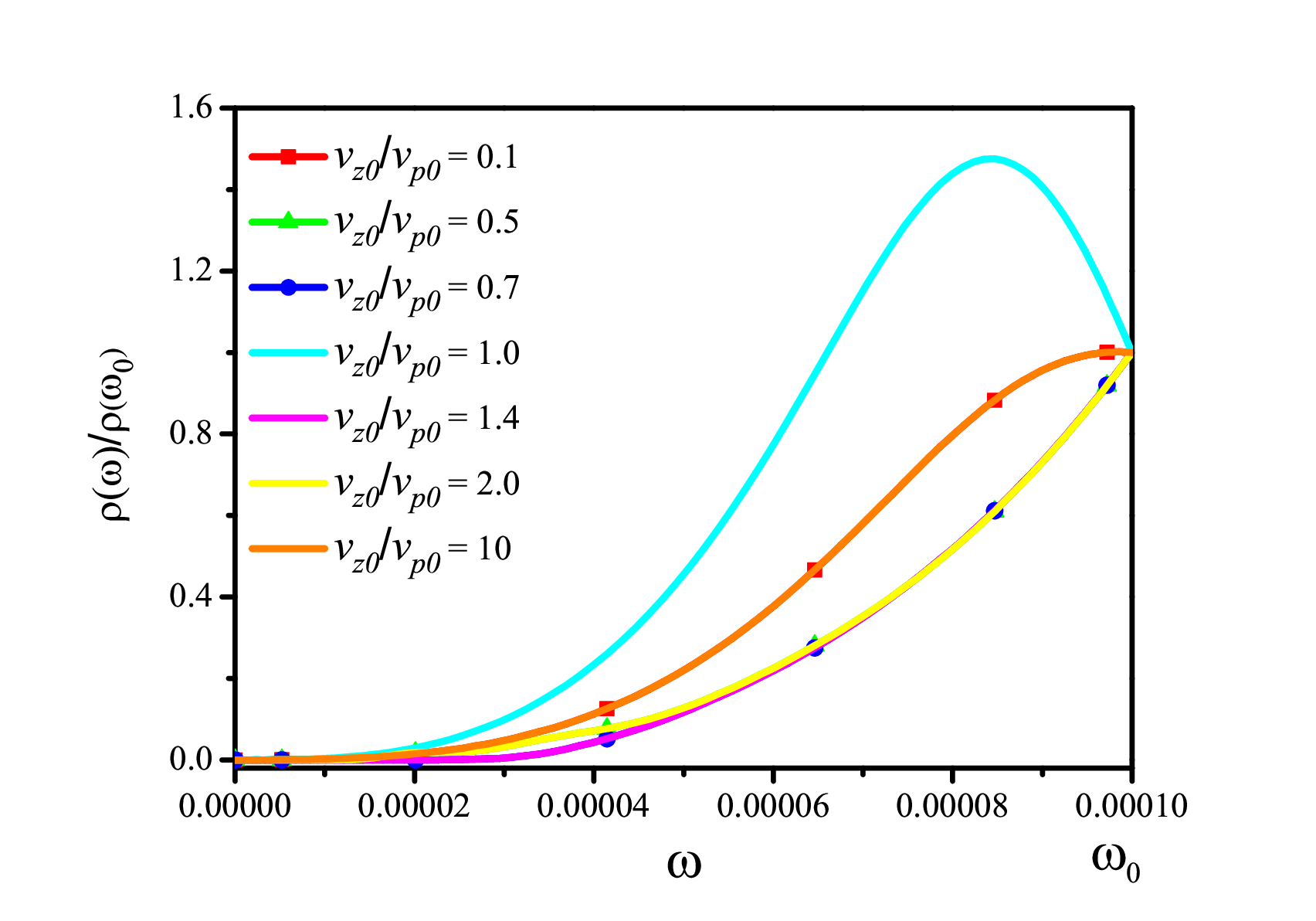}\\
\vspace{-0.25cm}
\caption{(Color online) Behavior of the DOS around FP-3 for
several representative values of fermion velocities ($v_{z0}/v_{p0}$).
The qualitative tendencies of DOS around FP-4,6,7,8 are analogous to that around FP-3 and hence
not shown hereby.}
\label{DOS-FP-3}
\end{figure}

Based on the free DOS, we then move to the situation involving the interactions
and disorder scatterings. Taking into account the corrections from the interplay between fermionic interactions and
disorders, and paralleling the analogous steps, we arrive at the renormalized DOS
\begin{widetext}
\begin{eqnarray}
\frac{\rho_{\mathrm{int}}(\omega)}{\Lambda_0k_F}
&=&\frac{N }{2\pi^2}
\int_0^{l_c}\int_0^\pi
\frac{e^{-2l}dld\theta}{\sqrt{v_z^2(l)\sin^2\theta+v_p^2(l)\cos^2\theta}}
\delta\Bigg(e^{-l}-\frac{|\omega|}
{\sqrt{v_z^2(l)\sin^2\theta+v_p^2(l)\cos^2\theta}}\Bigg).\label{rho_int_0}
\end{eqnarray}
\end{widetext}
Hereby, the $l_c$ denotes the critical energy scale at a certain fixed point, beyond which the
potential instability is induced and the RG approach is invalid.
In contrast to the free case, the fermion velocities, in this circumstance, are energy-dependent and
coupled with the other parameters via the RG equations~(\ref{Eq_v_RG})-(\ref{Eq_Delta_RG}).
This can be expected to bring a significant impact on the DOS.

Armed with the general expression of DOS~(\ref{rho_int_0}) and
the RG equations~(\ref{Eq_v_RG})-(\ref{Eq_Delta_RG}), we can now delve into a detailed
examination of the critical energy-dependent behavior of the DOS as
approaching all eight distinct types of fixed points shown in Figs.~\ref{class-FP-1-2}-\ref{class-FP-3-8}.

Before considering the interaction cases, let us examine the DOS of free case around the FP-1,
as illustrated in Fig.\ref{DOS-FP-1}. In this situation, the DOS exhibits a linear dependence on frequency $\rho_0(\omega)\propto|\omega|$,
and possesses the complete symmetry between $\omega>0$ and $\omega<0$, which is well consistent with Eq.~(\ref{rho_int_0}).
Apparently, the DOS precisely vanishes at the Fermi surface, and its the basic tendency is insusceptible to the
initial values of fermion velocities.

\begin{figure}[htpb]
\centering
\subfigure[]{\includegraphics[width=3.5in]{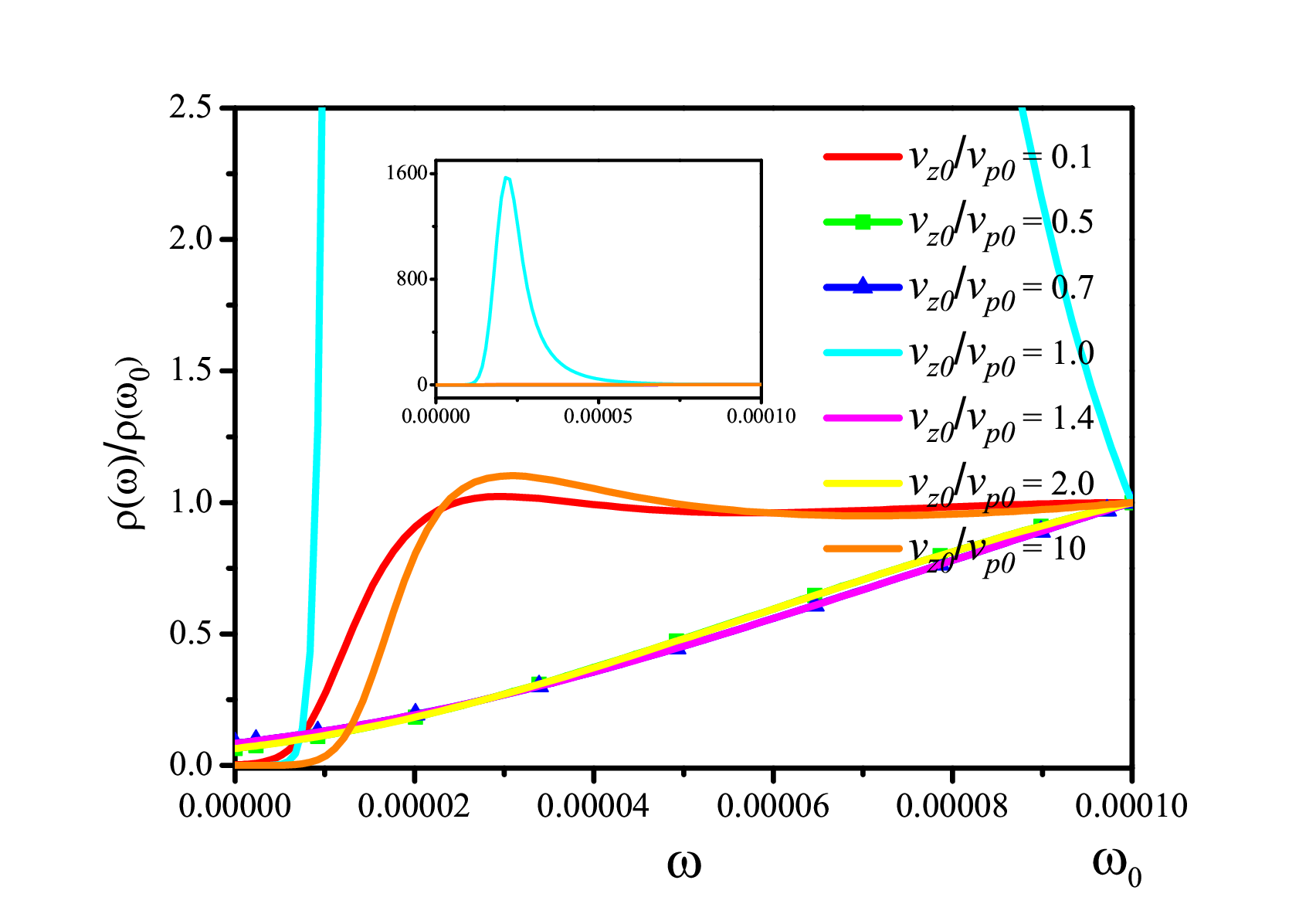}}\vspace{-0.5cm} \\
\subfigure[]{\includegraphics[width=3.5in]{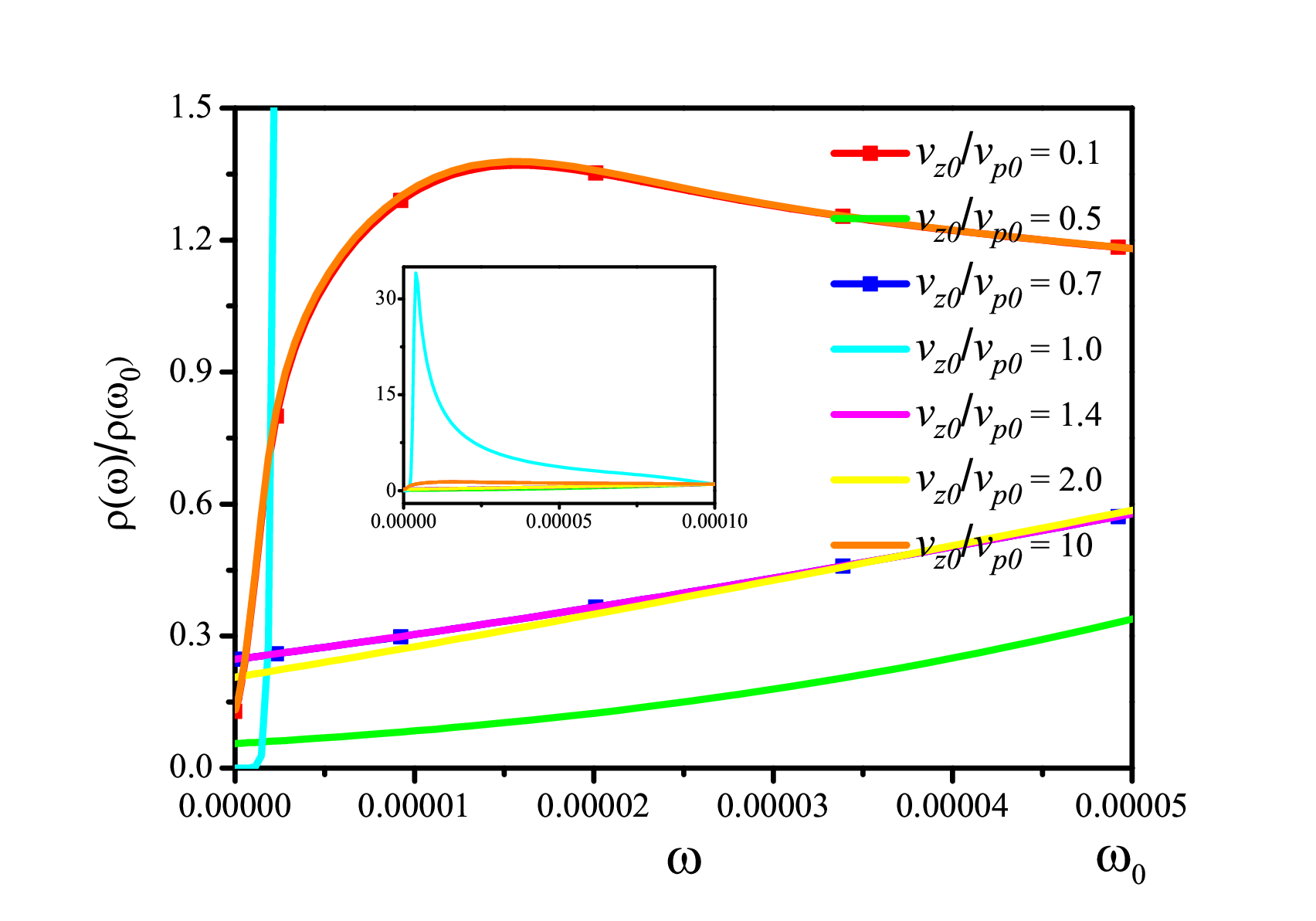}}\vspace{-0.5cm} \\
\subfigure[]{\includegraphics[width=3.5in]{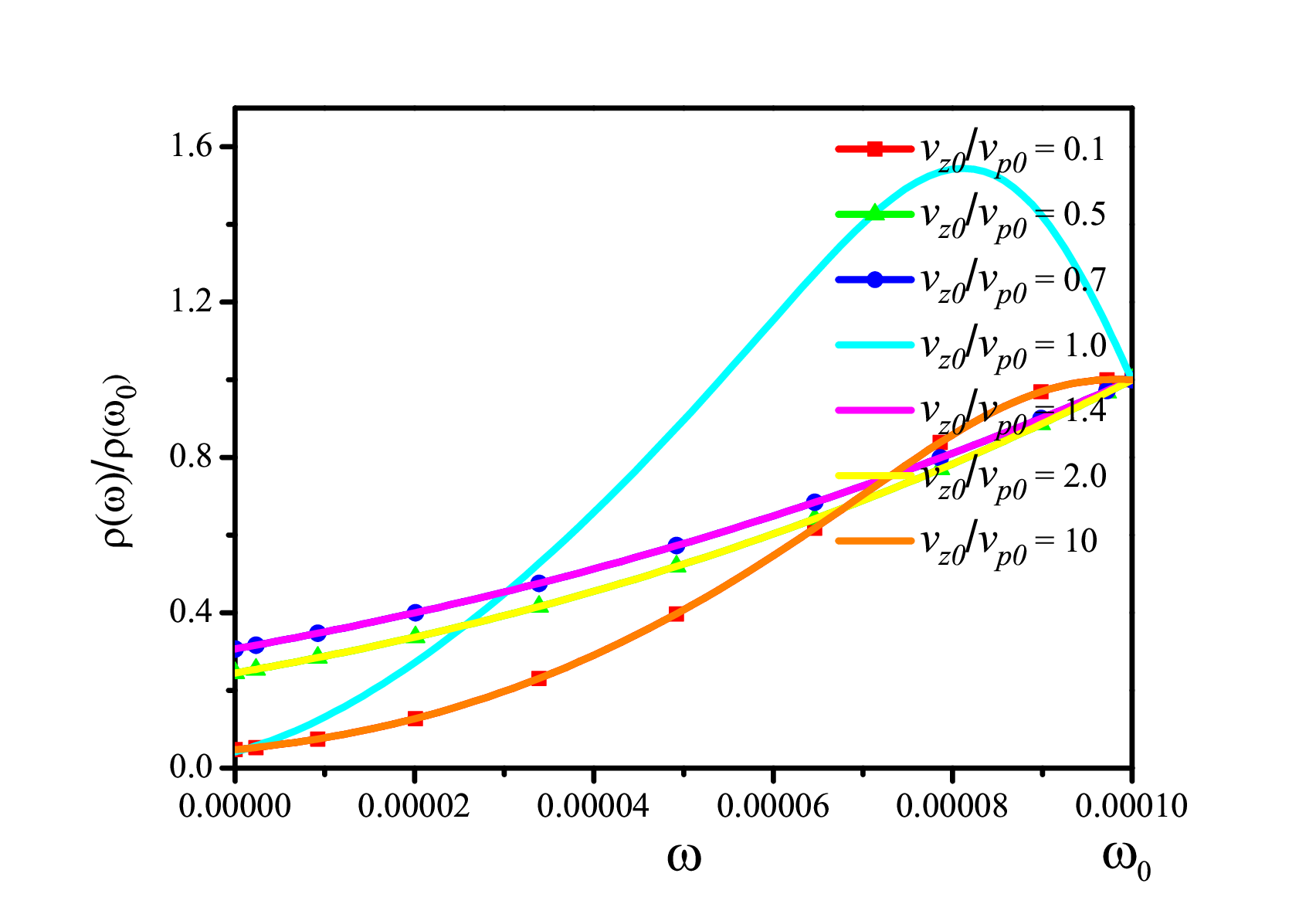}}\\
\vspace{-0.19cm}
\caption{(Color online) Behavior of the DOS around FP-5 with taking (a) $\Delta_2\neq0$ and
$\lambda_0=10^{-3}, 10^{-4}$, and (b) with $\Delta_{30}=10^{-5}$, as well as (c) $\Delta_{30}=10^{-7}$, for several
representative values of fermion velocities ($v_{z0}/v_{p0}$). Inset: the enlarged regime for $v_{z0}/v_{p0}=1$.}
\label{DOS-FP-5}
\end{figure}

In addition, we move our attention to FP-2 in the presence of the fermion-fermion interactions.
Numerically analysis with the help of RG equations yields the basic results, as presented in Fig.~\ref{DOS-FP-2}.
On one hand, we learn from Fig.~\ref{DOS-FP-2}(a) that the fermionic interactions can
coax the DOS to attain a nonzero finite value at $\omega=0$, which provides a qualitative
departure from the free case (FP-1) although the symmetry for
$\omega > 0$ and $\omega < 0$ is still preserved. Besides, we notice that such finite value is sensitive to the initial value of
Fermi velocities ($v_{z0}/v_{p0}$), which obtains the optimum value at $v_{z0}/v_{p0}\approx0.7,1.4$ and
minimum value in the isotropic situation, as depicted in Fig.~\ref{DOS-FP-2}(b)(For specific initial values of fermion velocities, such as $v_{z0}/v_{p0}=1$, a critical energy denoted as $\omega_c$ is identified, as illustrated in Fig.~\ref{DOS-FP-2}(b), which separates the evolution of DOS into two distinct regimes. The critical values ($\omega_c$) associated with other fermion velocities in Fig.~\ref{DOS-FP-3} and Fig.~\ref{DOS-FP-5} are defined in a manner analogous to that of Fig.~\ref{DOS-FP-2}(b). For the sake of simplicity, they are not explicitly labeled in Fig.~\ref{DOS-FP-3} and Fig.~\ref{DOS-FP-5}).
On the other hand, one can find that the DOS gradually increases with increasing the $\omega$ for
anisotropic fermion velocities. However, in the isotropic case with $v_{z0}/v_{p0}=1$,
Fig.~\ref{DOS-FP-2}(b) shows that the DOS increases when $\omega<\omega_c\approx0.000087$ and decreases beyond $\omega_c$.
We argue that this behavior can be ascribed to the competition between fermionic interactions and
thermal fluctuations, namely the thermal fluctuations are subordinate to the fermionic
interactions at $\omega>\omega_c$.

Furthermore, paralleling the analogous analysis to both FP-1 and FP-2,
we find that the DOS around the FP-3, 4, 6, 7, 8 share the similarly qualitative behavior
as illustrated in Fig.~\ref{DOS-FP-3}.  For simplicity, let us consider the FP-3 as an instance.
Under this circumstance, the interplay between the fermionic interactions and
the disorder scatterings leads to a quantitatively different DOS compared to FP-2, which
reduces to the value of DOS around FP-1 at $\omega\rightarrow0$. This indicates that the contribution of
the fermionic interactions is subordinate to disorder scatterings in the
low-energy regime ($\omega\rightarrow0$). However, as $\omega$ increases, the DOS exhibits a more similar behavior
to its FP-2 counterpart, namely gradually increasing with the rise of $\omega$ under
the influence of anisotropic fermion velocities. In comparison, when the fermion velocities are
isotropic, the DOS initially increases and then decreases below the critical frequency $\omega_c$.
This implies that the contributions of fermion interactions and disorder scatterings are subordinate to thermal fluctuations
at $\omega<\omega_c$ and vice versa at $\omega>\omega_c$. Besides, a bigger anisotropy of fermion velocities
is helpful to increase the DOS.

\begin{figure}[htpb]
\centering
\includegraphics[width=3.5in]{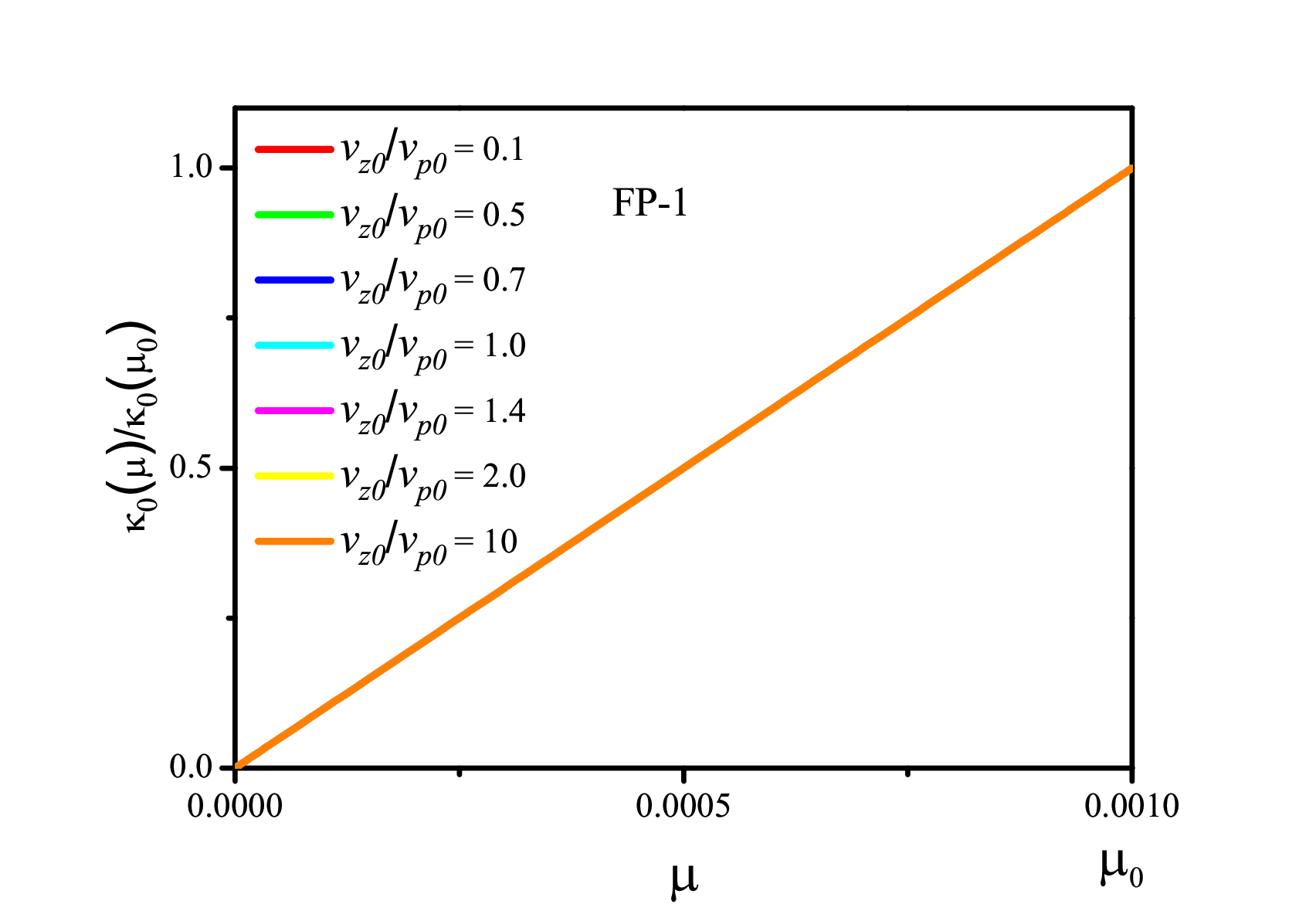}\\
\vspace{-0.15cm}
\caption{(Color online) Behavior of the compressibility $\kappa(\mu)$ around FP-1
(free case) for several representative values of fermion velocities ($v_{z0}/v_{p0}$)
with $\mu_0=0.001$.}
\label{kappa-FP-1}
\end{figure}

\begin{figure}[htpb]
\centering
\includegraphics[width=3.5in]{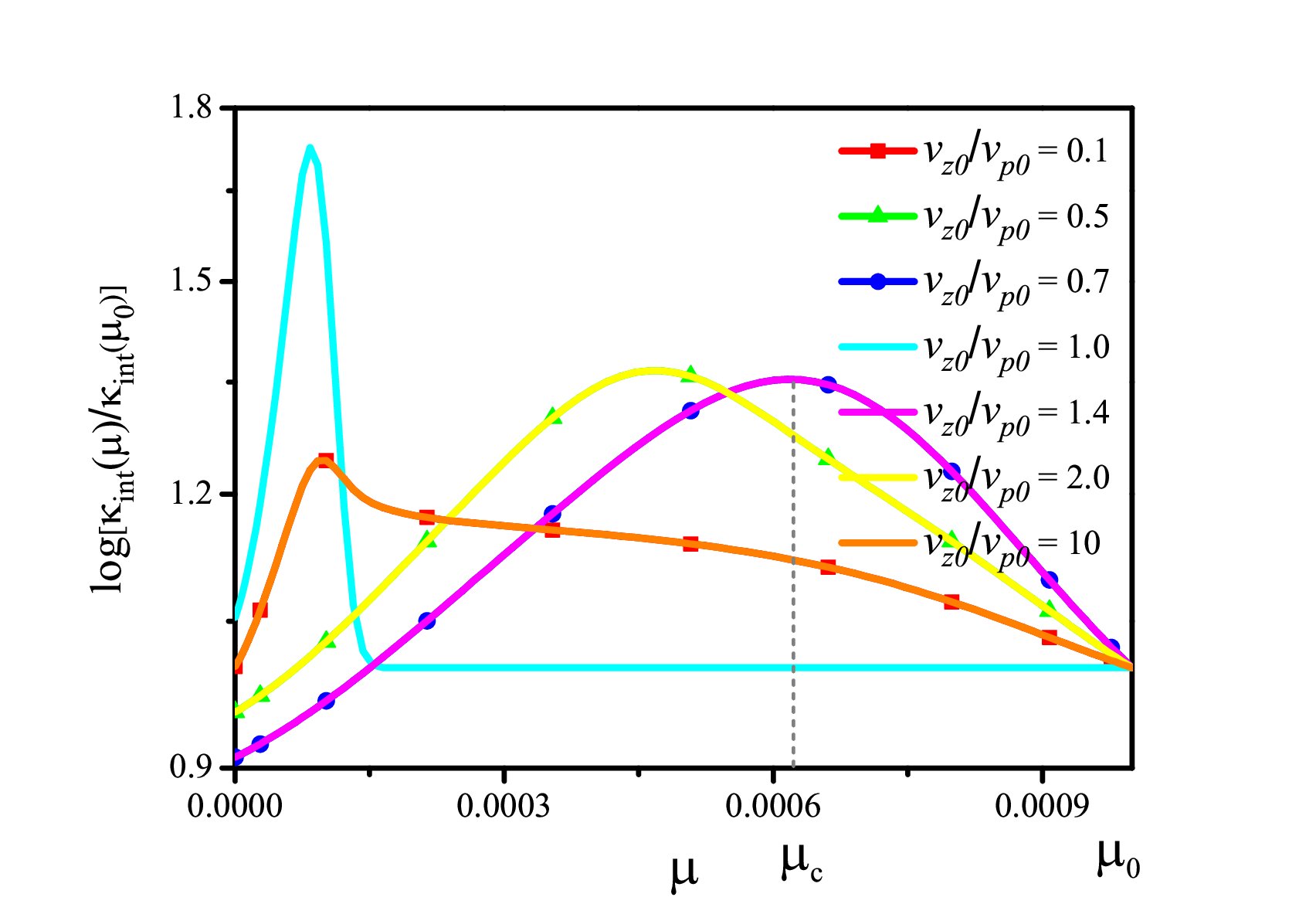}\\ 
\vspace{-0.15cm}
\caption{(Color online) Behavior of the compressibility $\kappa(\mu)$ around FP-2 for
several representative values of fermion velocities ($v_{z0}/v_{p0}$). The qualitative tendencies
of compressibility around FP-3,4,6,7 are analogous to that around FP-2 and hence
not shown hereby.} 
\label{kappa-FP-2}  
\end{figure}

At last, turning our attention to the behavior of the DOS around FP-5 as shown in Fig.~\ref{DOS-FP-5},
we summarize the evolution of the DOS with frequency in the sole presence of $\Delta_2$ or $\Delta_3$
for several representative values of fermion velocities.
On the one hand, we notice that the fermion velocities play a more significant role in the low-energy regime.
In the presence of $\Delta_2$ alone, as illustrated in Fig.~\ref{DOS-FP-5}(a), the DOS vanishes at both strong anisotropy and isotropy of fermion velocities at $\omega=0$. However, weak anisotropy results in a finite value of DOS.
In contrast, when considering the single presence of $\Delta_3$, the results are sensitively dependent
on the initial value of disorder strength. Fig.~\ref{DOS-FP-5}(b) with a bigger $\Delta_3$
demonstrates that the DOS vanishes at the
isotropy of fermion velocities, while the anisotropy leads to a finite DOS at $\omega=0$. However,
Fig.~\ref{DOS-FP-5}(c) shows that the DOS can obtain a finite value at $\omega=0$ for a weaker $\Delta_3$.
As a consequence, the anisotropy of fermion velocities quantitatively determines the value of the DOS at $\omega = 0$ in the presence of $\Delta_2$ or a larger $\Delta_3$, whereas the presence of a weaker $\Delta_3$ does not quantitatively change the magnitude of the DOS at $\omega=0$. On the other hand, once the fermion velocities are isotropic or stronger anisotropic,
as depicted in Fig.~\ref{DOS-FP-5}(b)-(c), the DOS experiences an increase below the critical frequency $\omega_c$
but gradually decreases above it (except the isotropic case with a weaker $\Delta_3$, where it monotonously increases).
This implies that a bigger anisotropy of fermion velocities at $\omega<\omega_c$ tends to assist
the thermal fluctuations in suppressing both the fermionic interaction and disorder scatterings, while the thermal fluctuations
become subordinate to the interactions at higher frequencies.

To wrap up, as we approach the distinct kinds of FPs,
the competition among fermion-fermion interactions and disorder scatterings as well as thermal fluctuations
results in several distinct kinds of critical behavior of the DOS.

\subsection{Compressibility}\label{Subsection_kappa}

Subsequently, we shift our focus to the compressibility of quasiparticles,
which is a significant physical quantity to characterize the flexibility of a certain system.
Based on the original definition, the
compressibility is designated as $\kappa=\partial V/\partial F$, where $V$
and $F$ represent the volume and compressing force, respectively.
However, it is more convenient to denominate the compressibility as $\kappa=\partial n/\partial \mu$ in
the calculation of many-particle systems, where $n$ is the number of particles per unit volume
(or particle number density)~\cite{Schwabl2006Book,Altland2006Book}.

In order to derive the expression of compressibility for our theory, we need to introduce the chemical potential $\mu$
as an auxiliary variable and obtain the $\mu$-dependent DOS~\cite{Wang2014PRD}.
At a finite chemical potential $\mu$, the free fermionic propagator can be reformulated as
\begin{eqnarray}
G_0(\omega,\mathbf{k})
=
\frac{1}{-i\omega-\mu+v_z\delta k_z\Sigma_{03}+v_p\delta k_\perp \Sigma_{01}}.
\end{eqnarray}
This gives rise to the non-interacting DOS after paralleling the method in Sec.~\ref{Subsection_DOS}
\begin{eqnarray}
\rho_0(\omega)
&=&
N \int\frac{d^3\mathbf{k}}{(2\pi)^3}\mathrm{Tr}\left[\mathcal{A}(\omega,\mathbf{k})\right]\nonumber\\
&=&
\frac{N k_F |\omega+\mu|}{2\pi^2}\alpha(v_z,v_p),\label{rho_0}
\end{eqnarray}
which reduces to a constant proportional to $\mu$ as $\omega$ vanishes.
To proceed, we take into account the influences of the interplay between fermion-fermion interactions and
disorder scatterings. Carrying out the long but straightforward calculations yields the renormalized DOS
as follows
\begin{widetext}
\begin{small}
\begin{numcases}
{\rho_{\mathrm{int}}(\omega+\mu) = }
\frac{N k_F}{2\pi^2}\int_0^{\Lambda_0}\int_0^\pi
\frac{kdkd\theta}{\sqrt{v_z^2(k)\sin^2\theta+v_p^2(k)\cos^2\theta}}
\delta\Big\{k-\frac{\omega+\mu}{\sqrt{v_z^2(k)\sin^2\theta+v_p^2(k)\cos^2\theta}}\Big\}
,& $\mu>0$ \nonumber\\
\frac{N k_F}{2\pi^2}\int_0^{\Lambda_0}\int_0^\pi
\frac{kdkd\theta}{\sqrt{v_z^2(k)\sin^2\theta+v_p^2(k)\cos^2\theta}}
\delta\Big\{k+\frac{\omega+\mu}{\sqrt{v_z^2(k)\sin^2\theta+v_p^2(k)\cos^2\theta}}\Big\}
,& $\mu<0$\label{rho_int}
\end{numcases}
\end{small}
\end{widetext}
where $N$ specifies the fermion flavor and $k_F$ denotes the Fermi momentum, respectively.

\begin{figure}[htpb]
\centering
\includegraphics[width=3.5in]{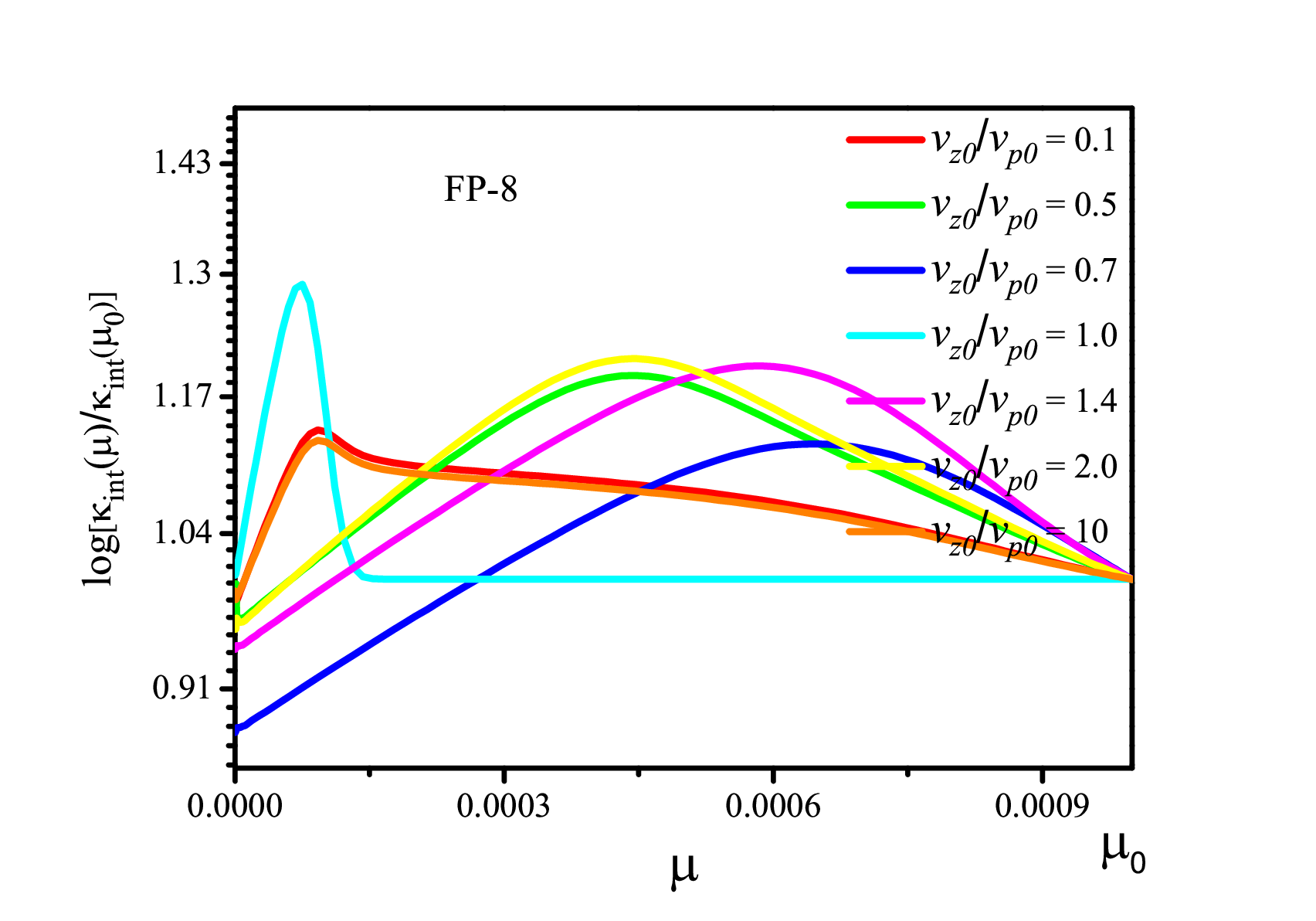}\\
\vspace{-0.2cm}
\caption{(Color online) Behavior of the compressibility $\kappa(\mu)$ around FP-8 for
several representative values of fermion velocities ($v_{z0}/v_{p0}$). The qualitative tendency
of compressibility around FP-5 is analogous to that around FP-8 and hence not shown hereby.}
\label{kappa-FP-8}
\end{figure}

With the $\mu-$dependent DOS~(\ref{rho_0})-(\ref{rho_int}) in hand, we are now in a proper position to
derive the expression of the compressibility. At first, let us calculate the $\mu-$dependence of particle number $n$.
For the non-interacting situation, utilizing Eq.~(\ref{rho_0}) leads to
\begin{eqnarray}
n_0(\mu)
&=&
\int_{-\mu}^0d\omega\rho_0(\omega)
=
\int_{-\mu}^0d\omega\frac{N k_F (\omega+\mu)}{2 \pi^2}\alpha(v_z,v_p)\nonumber\\
&=&
\frac{N k_F \mu^2}{4 \pi^2}\alpha(v_z,v_p),
\end{eqnarray}
with which, the free-case compressibility can be derived
\begin{eqnarray}
\frac{\kappa_0(\mu)}{\Lambda_0 k_F}
=
\frac{1}{\Lambda_0 k_F} \frac{\partial }{\partial\mu} n_0(\mu)
=
\frac{N \mu}{2\pi^2}\alpha(v_z,v_p).\label{kappa_0}
\end{eqnarray}
Here, it is worth highlighting that the above expressions are applicable to both $\mu>0$ and $\mu<0$.
In addition, we rescale the chemical potential by $\Lambda_0$,
indicating $\mu \to \mu/\Lambda_0\in (-1,1)$.
Subsequently, employing Eq.~(\ref{rho_int}) and following similar steps for derivation of
interacting DOS, we obtain the compressibility influenced by the fermionic interaction and disorder scatterings,
\begin{widetext}
\begin{small}
\begin{numcases}
{\frac{\kappa_{\mathrm{int}}(\mu)}{\Lambda_0 k_F} = }
\frac{N k_F}{2\pi^2}\int_0^{l_c}\int_0^\pi
\frac{e^{-2l}dld\theta}{\sqrt{v_z^2(l)\sin^2\theta+v_p^2(l)\cos^2\theta}}
\delta\Big\{e^{-l}-\frac{\mu}{\sqrt{v_z^2(l)\sin^2\theta+v_p^2(l)\cos^2\theta}}\Big\}
, & $\mu>0$ \nonumber\\
-\frac{N k_F}{2\pi^2}\int_0^{l_c}\int_0^\pi
\frac{e^{-2l}dld\theta}{\sqrt{v_z^2(l)\sin^2\theta+v_p^2(l)\cos^2\theta}}
\delta\Big\{e^{-l}+\frac{\mu}{\sqrt{v_z^2(l)\sin^2\theta+v_p^2(l)\cos^2\theta}}\Big\}
, & $\mu<0$\label{kappa_int}
\end{numcases}
\end{small}
\end{widetext}
where the variable $l_c$ serves as the critical energy scale at a certain fixed point illustrated in
Fig.~\ref{class-FP-1-2}.

Subsequently, armed with the analytical expressions~(\ref{kappa_0})-(\ref{kappa_int}),
we can now examine the behavior of the compressibility dubbed $\kappa$ as the system approaches the
distinct kinds of FPs illustrated in Fig.~\ref{class-FP-3-8}.

At the outset, we consider the evolution of $\kappa$ around the FP-1.
It is of particular interest to emphasize that the compressibility exhibits the symmetry between $\mu<0$ and $\mu>0$ as well,
which is similar to that of DOS shown in Fig.~\ref{DOS-FP-1}. To simplify our analysis,
we hereafter only focus on the $\mu>0$ situation. Fig.~\ref{kappa-FP-1} showcases the behavior of $\kappa$ for
the free case. It can be find that $\kappa$ is proportional to the chemical potential $\mu$
(i.e., $\kappa(\mu)\propto\mu$), indicating the vanish of compressibility at $\mu=0$.
Besides, it can be learned from Fig.~\ref{kappa-FP-1} that the tendency of $\kappa$ is insensitive to
the initial anisotropies of fermion velocities.

\begin{figure}[htpb]
\centering
\includegraphics[width=3.5in]{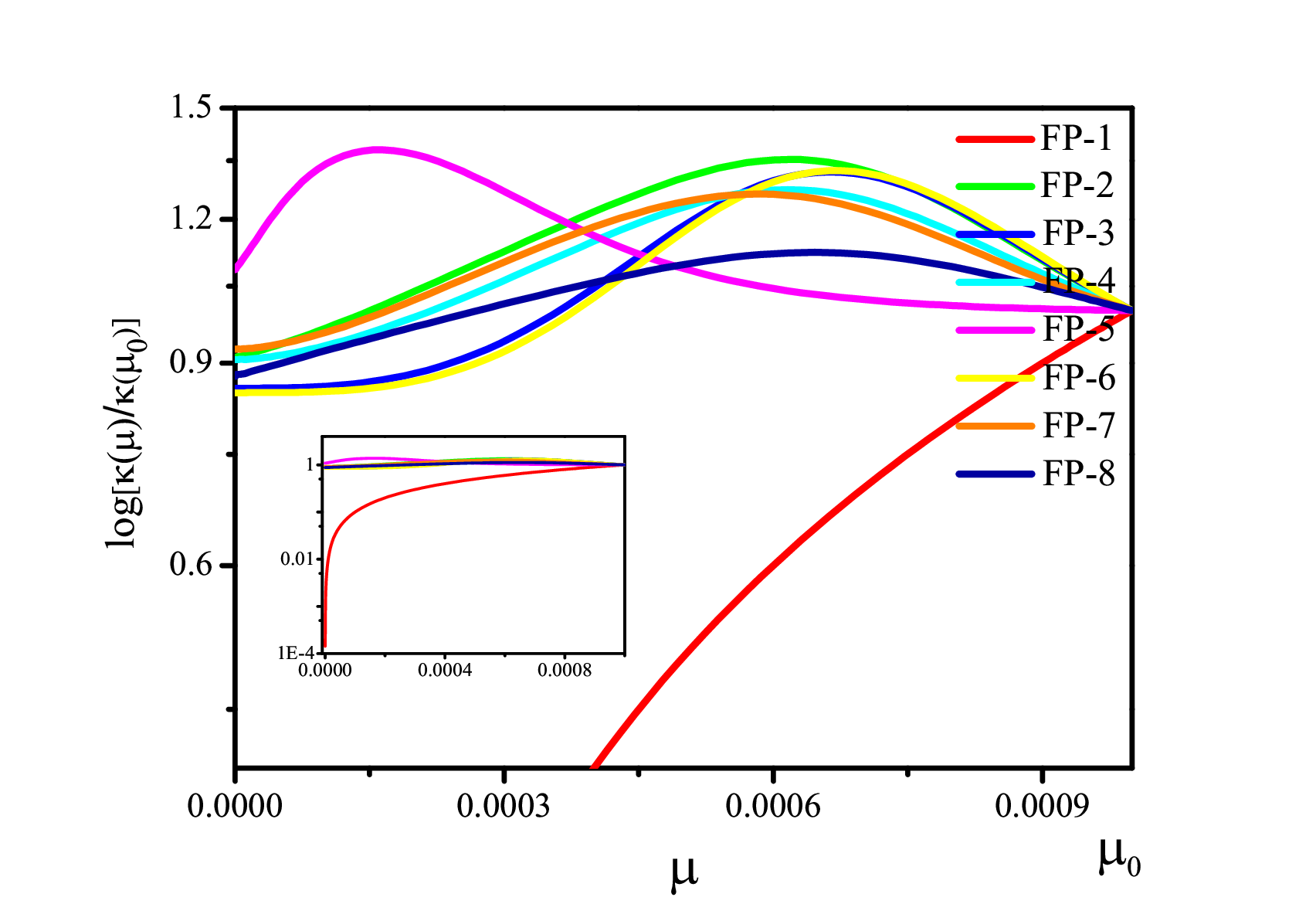}\\
\vspace{-0.2cm}
\caption{(Color online) Comparisons of compressibility $\kappa(\mu)$ around all eight
distinct types of FPs depicted in Fig.~\ref{class-FP-1-2} and Fig.~\ref{class-FP-3-8} with a fixed ratio of fermion
velocities $v_{z0}/v_{p0}=0.7$.}
\label{kappa-FP-1-8}
\end{figure}

To proceed, we will examine the behavior of $\kappa$ around the other FPs.
Given that FP-2, FP-3, FP-4, FP-6, and FP-7 share the analogously basic results, we hereby select the
FP-2 as an instance to present the related properties, as depicted in Fig.~\ref{kappa-FP-2} (For specific initial values of fermion velocities, such as $v_{z0}/v_{p0}=1$, a critical value of chemical potential denoted as $\mu_c$ is identified, as illustrated in Fig.~\ref{kappa-FP-2}, which separates the evolution of compressibility into two distinct regimes. The critical values ($\mu_c$) associated with other fermion velocities in Fig.~\ref{kappa-FP-8} and Fig.~\ref{kappa-FP-1-8} are defined in a manner analogous to that of Fig.~\ref{kappa-FP-2}. For the sake of simplicity, they are not explicitly labeled in Fig.~\ref{kappa-FP-8} and Fig.~\ref{kappa-FP-1-8}).  Observing Fig.~\ref{kappa-FP-2}, we note that the compressibility, below the critical value $\mu_c$, initially increases as $\mu$ rises,
but gradually decreases while $\mu$ goes beyond $\mu_c$. Again, we realize that these results are insensitive to
the starting anisotropies of fermion velocities. This may suggest certain intricate competition among the fermionic interactions,
disorder scatterings and the thermal fluctuations. In other words, the contributions from the interactions and disorders are
subordinate to the thermal fluctuations at $\mu<\mu_c$, but instead the interactions and disorders dominate at $\mu>\mu_c$.
In addition, although the basic results are analogous, the specific value of the compressibility at $\mu=0$ is of close
relevance to the anisotropy of fermion velocities as illustrated in Fig.~\ref{kappa-FP-2}.
On one hand, an isotropic or a stronger anisotropy of fermion velocities ($v_{z0}/v_{p0}$ or $v_{p0}/v_{z0}$) yields a
bigger $\kappa(0)$. On the other hand, we realize that the anisotropy of fermion velocities
affects the critical chemical potential ($\mu_c$) and the associated compressibility as well.
Specifically, a bigger anisotropy of fermion velocities is prefer to suppress both $\mu_c$
and $\kappa(\mu_c)$. In particular, the isotropy of fermion velocities leads to the smallest $\mu_c$.
As a consequence, the ratio of fermion velocities plays a key role in determining the
behavior of $\kappa$ around FP-2, FP-3, FP-4, FP-6, and FP-7.
With an adequate strong anisotropy of fermion velocities, the combined effects of interactions
and disorders prevail over thermal fluctuations.

Next, we turn our attention to the properties of $\kappa$ around FP-5 and FP-8, both of which exhibit
qualitatively analogous tendencies.
Let us take the FP-8 for an instance, around which the related results are illustrated in Fig.~\ref{kappa-FP-8}.
In comparison to the FPs discussed above, the compressibility $\kappa(\mu)$ presents several unique properties
under the competition between the interactions and the thermal fluctuations.
At the first sight of Fig.~\ref{kappa-FP-8}, we notice that
$\kappa(\mu)$ gradually rises and arrives at a certain saturated value as $\mu$ increases within the region of $\mu<\mu_c$.
Conversely, the compressibility progressively decreases while $\mu$ is beyond $\mu_c$.
Additionally, the anisotropy of fermion velocities is a crucial ingredient to determine
the critical chemical potential ($\mu_c$) and the corresponding compressibility.
In this situation, one can find the basic results are analogous to those of FP-2, as depicted in Fig.~\ref{kappa-FP-2}.

Furthermore, Fig.~\ref{kappa-FP-1-8} summarizes the behavior of compressibility in the vicinity of
all eight distinct types of FPs. It is evident that the basic tendencies of $\mu$-dependent evolutions of $\kappa(\mu)$ are analogous, specifically an increase below the critical chemical potential ($\mu_c$) and a decrease beyond $\mu_c$. The presence of fermionic interactions and disorder scatterings leads to a significant enhancement in compressibility owing to the intricate interplay between these different kinds of physical ingredients.

\begin{figure}[htpb]
\centering
\subfigure[]{\includegraphics[width=3.5in]{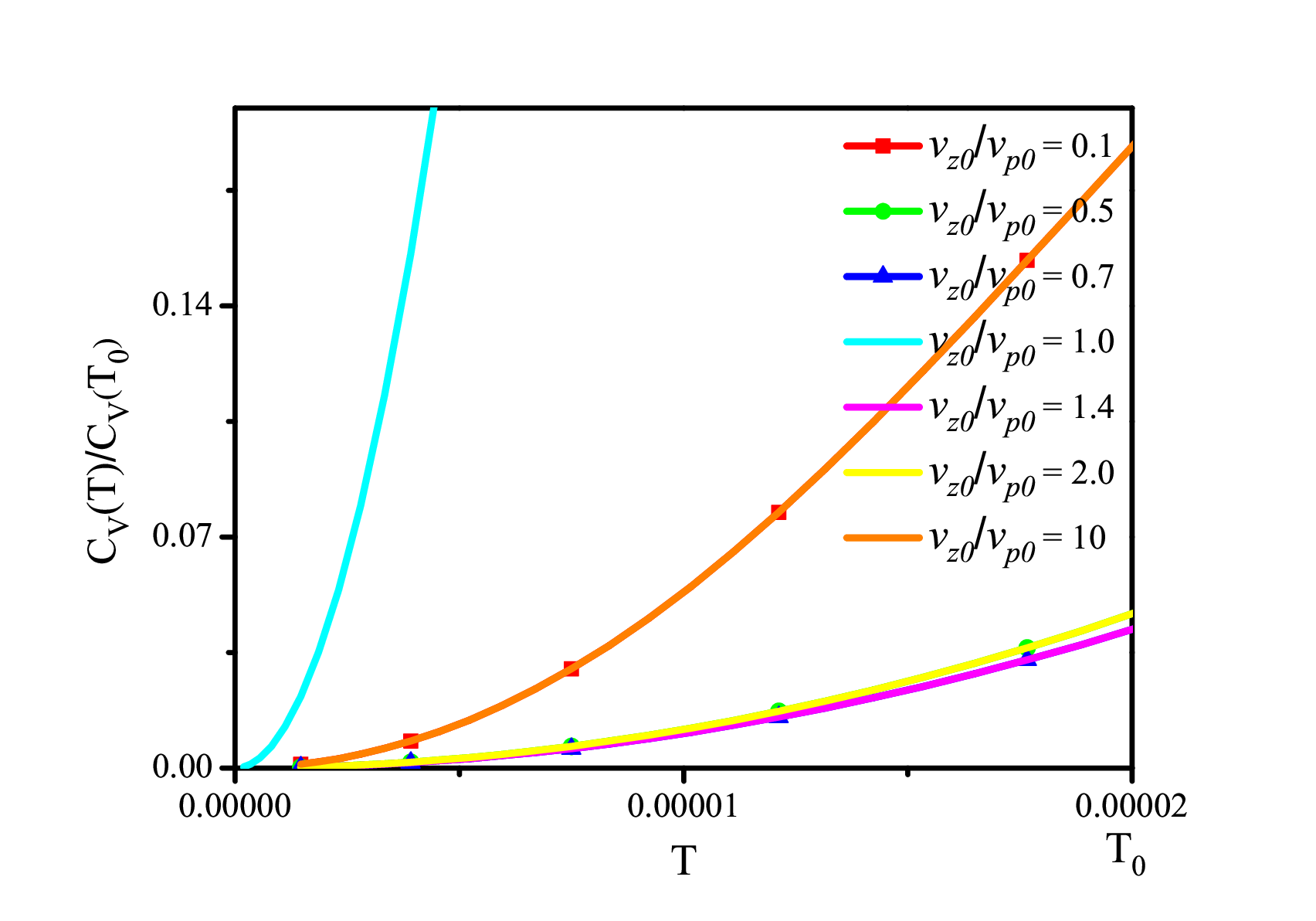}} \\ \vspace{-0.5cm}
\subfigure[]{\includegraphics[width=3.5in]{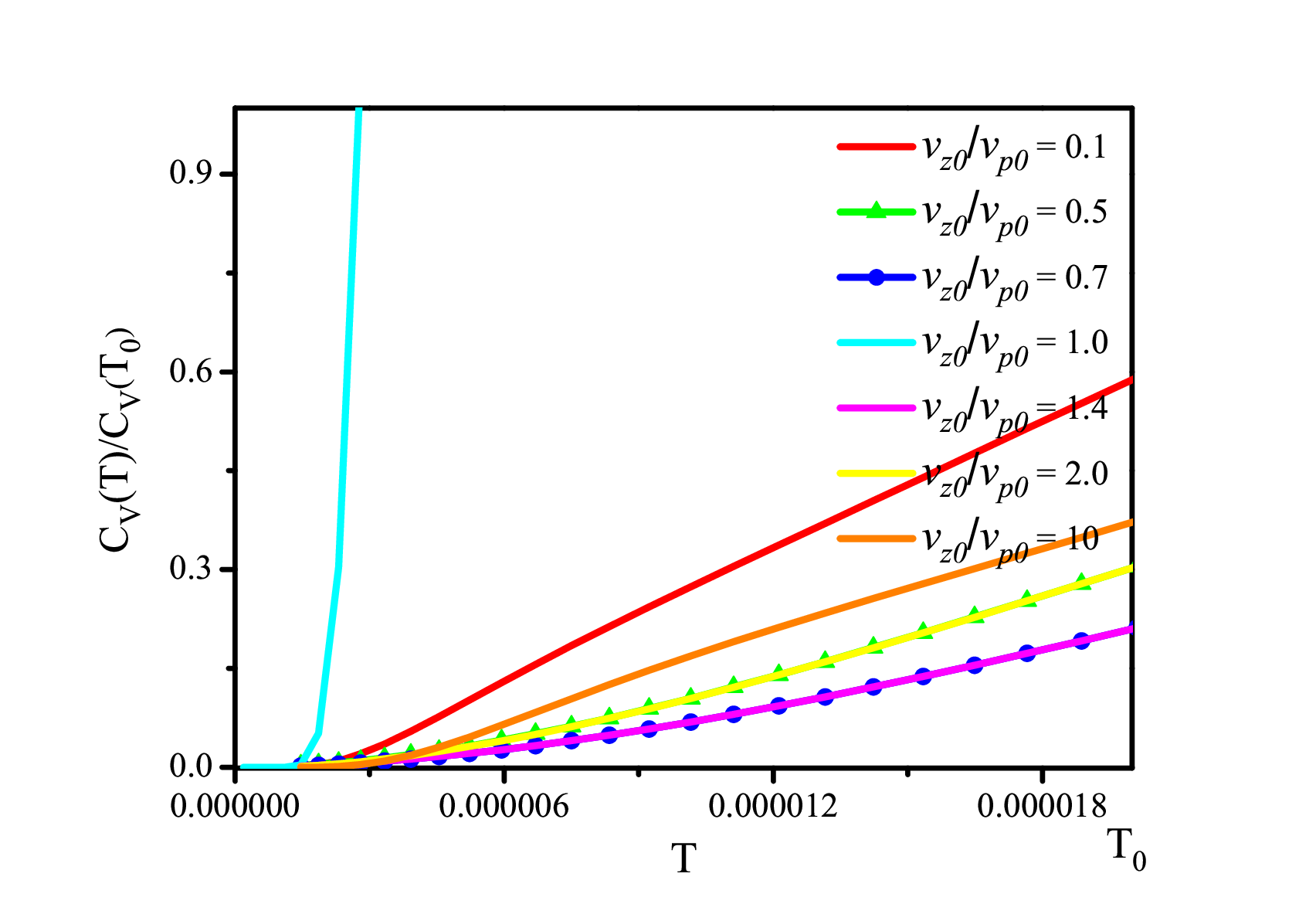}}\\
\vspace{-0.25cm}
\caption{(Color online) Behavior of the $C_V(T)$ around (a) FP-1
and (b) FP-5 for several representative values of fermion velocities ($v_{z0}/v_{p0}$).
The qualitative tendencies of $C_V(T)$ around FP-2,3,4,6,7,8 are analogous to
that around FP-1 and hence not shown hereby.} 
\label{CV-FP-1-5}  
\end{figure}

\renewcommand\arraystretch{2.0}
\begin{table*}[htbp]
\caption{Six distinct sorts of potential candidate states in the superconducting dome of
nodal-line supercionductors:  1) SC-I ($is$-pairing), 2) SC-II ($ig$-pairing), 3)
SC-III ($id_{x^2-y^2}$-pairing), 4) SC-IV ($id_{xy}$-pairing), 5)
SC-V ($id_{xz}$-pairing), and 6) SC-VI ($id_{yz}$-pairing)~\cite{Moon2017PRB}.
Hereby, the matrix $\mathcal{M}_i$ characterizes the coupling between fermions and source terms, which
are related to the certain symmetry breaking.}
\label{table_source} \vspace{0.3cm} 
\begin{tabular}{p{3.5cm}<{\centering} p{2.1cm}<{\centering} p{2.35cm}<{\centering}
p{2.0cm}<{\centering} p{2.35cm}<{\centering} p{2.0cm}<{\centering} p{2.0cm}<{\centering}}
\hline \hline Order parameter ($g_i$)  & SC-I & SC-II & SC-III
& SC-IV
& SC-V & SC-VI \\
\hline Coupling matrices ($\mathcal{M}_i$) & $\gamma_{02}$ &
$\frac{1}{4}\sin(4\theta_{\mathbf{k}})\gamma_{02}$ &
$\cos(2\theta_{\mathbf{k}})\gamma_{02} $ &
$\frac{1}{2}\sin(2\theta_{\mathbf{k}})\gamma_{02} $ &
$\cos(\theta_{\mathbf{k}})\gamma_{02} $
& $\sin(\theta_{\mathbf{k}})\gamma_{02}$ \\
\hline \hline
\end{tabular}
\end{table*}

\subsection{Specific heat}\label{Subsection_C_V}

At last, let us study the specific heat of quasiparticles.
To simplify our analysis, we employ the ultraviolet cutoff ($\Lambda_0$)
to rescale the related quantities, namely $T, k, \mu\to T/\Lambda_0, k/\Lambda_0, \mu/\Lambda_0$.
To proceed, the partition function for the fermionic quasiparticles can be written as~\cite{Kapusta1994Book}
\begin{eqnarray}
Z
&=&
\prod_{n,\mathbf{k},\alpha}\int\mathcal{D}[-i\psi_{\alpha,n}^\dag(\mathbf{k})]
\mathcal{D}[\psi_{\alpha,n}(\mathbf{k})]
e^S,
\end{eqnarray}
where the action $S$ takes the form of
\begin{eqnarray}
S&=&
\sum_{n,\mathbf{k}}[-i\psi_{\alpha,n}^\dag(\mathbf{k})]D_{\alpha,\rho}
[\psi_{\rho,n}(\mathbf{k})],\nonumber\\
D
&=&\beta\left[\omega_n
+i(v_z\delta k_z\Sigma_{03} +v_p\delta k_\perp\Sigma_{01})\right],
\end{eqnarray}
with the inverse of temperature $\beta=1/T$.
Then, employing the formula
\begin{eqnarray}
\int \mathcal{D}[\eta^\dag]\mathcal{D}[\eta]e^{\eta^\dag D \eta}
=
\det D,
\end{eqnarray}
to integrate out the functions of Grassmann variables yields
\begin{eqnarray}
\ln Z
&=&
\sum_{n, \mathbf{k}}
\ln\big\{\beta^2[\omega_n^2+\epsilon^2(\mathbf{k})]^2\big\},
\end{eqnarray}
where $\epsilon(\mathbf{k})=v_z^2\delta k_z^2+v_p^2\delta k_\perp^2$. Considering the following identities,
\begin{eqnarray}
&&\int_1^{\beta^2\epsilon^2(\mathbf{k})}
\frac{d\theta^2}{\theta^2+(2n+1)^2\pi^2}
+
\ln [1+(2n+1)^2\pi^2]\nonumber\\
&&=
\ln\left[(2n+1)^2\pi^2+\beta^2\epsilon^2(\mathbf{k})\right],\\
&&\sum_{-\infty}^\infty\frac{1}{(n-x)(n-y)}
=
\frac{\pi[\cot(\pi x)-\cot(\pi y)]}{y-x},
\end{eqnarray}
we finally arrive at
\begin{eqnarray}
\ln Z
&=&
\sum_k 2 \left[2\ln(1+e^{-\beta\epsilon(\mathbf{k})})+\beta\epsilon(\mathbf{k})\right],\nonumber\\
&=&
V\int \frac{d^3\mathbf{k}}{(2\pi)^3}\big[4\ln(1+e^{-\beta\epsilon(\mathbf{k})})+2\beta\epsilon(\mathbf{k})\big].
\end{eqnarray}

\begin{figure}[htpb]
\centering
\includegraphics[width=3.5in]{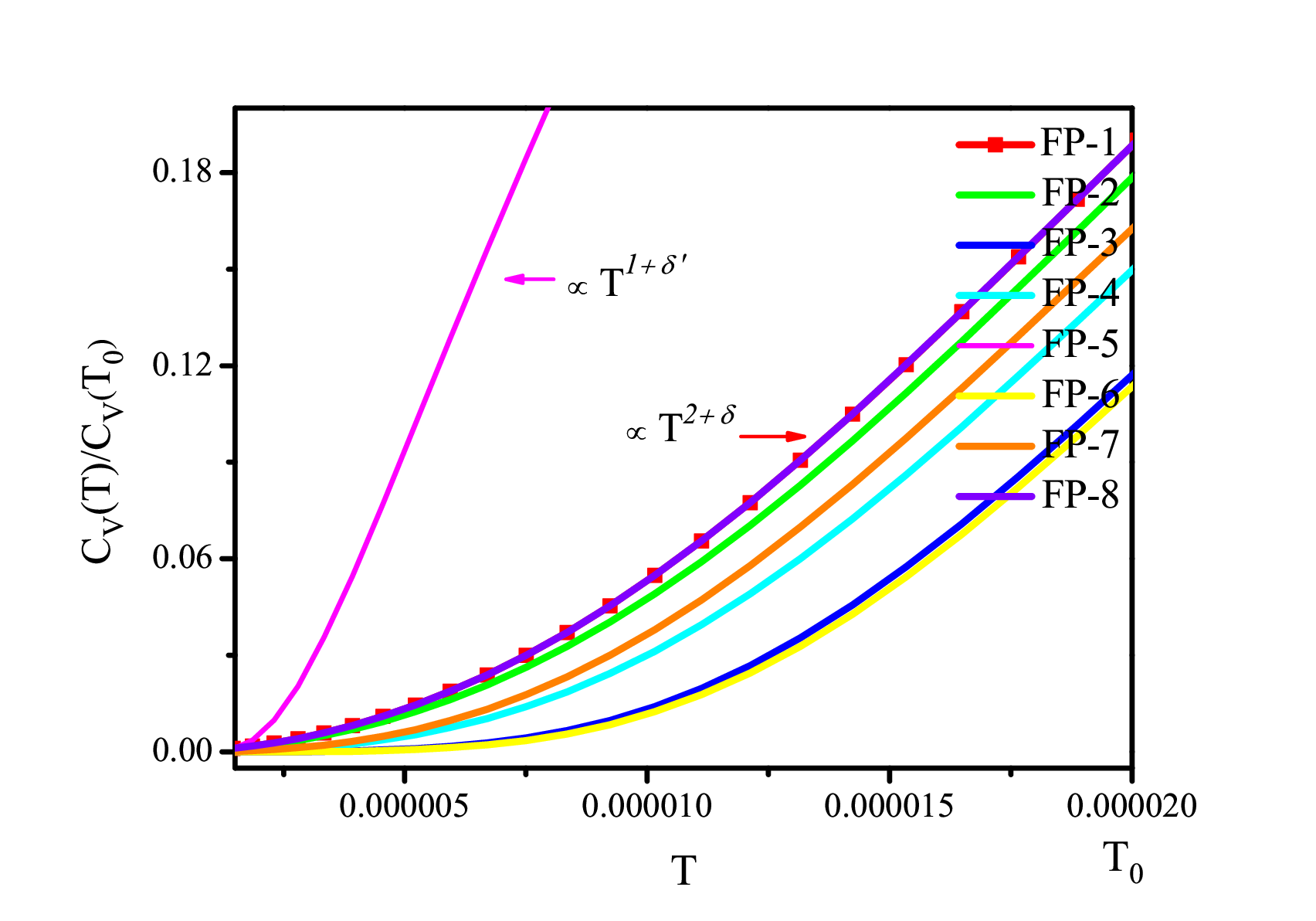}\\
\vspace{-0.2cm}
\caption{(Color online) Comparisons of specific heat $C_V(T)$ around all eight
distinct types of FPs depicted in Fig.~\ref{class-FP-1-2} and Fig.~\ref{class-FP-3-8} with a fixed ratio of fermion
velocities $v_{z0}/v_{p0}=0.1$. Hereby, the curves for FP-1 and FP-5 are approximately proportional to $T^{2+\delta}$
and $T^{1+\delta'}$ with $\delta,\delta'\ll1$, respectively.}
\label{CV-multiply-FPs}
\end{figure}

Moving forward, employing $f=F/V=-\ln Z/(\beta V)$ with $\beta=1/T$, the free energy density of fermions can be expressed as~\cite{Kapusta1994Book}
\begin{eqnarray}
f(T)
&=&
-\int\frac{d^3\mathbf{k}}{(2\pi)^3}
\left\{4T\ln\Big[1+\exp\big(-\frac{\epsilon(\mathbf{k})}{T}\big)\Big]+2\epsilon(\mathbf{k})\right\},\nonumber\\
\end{eqnarray}
where the zero-point energy (the second term in $\ln Z$) has
been discarded. Armed with the free energy density, we can easily obtain the specific heat~\cite{Altland2006Book},
\begin{eqnarray}
C_V=-T\frac{\partial^2 f(T)}{\partial T^2}.
\end{eqnarray}
After carrying out several calculations in the absence of fermion-fermion interactions
and disorder scatterings, we derive the specific heat for the free case,
\begin{widetext}
\begin{eqnarray}
\frac{C_V^0(T)}{\Lambda_0^2 k_F}
&=&
\frac{1}{\pi^2 T^2}
\int_0^1 \int_0^\pi dk d\theta
\left\{
\frac{ k^3(v_z^2\sin^2\theta+v_p^2\cos^2\theta)
\exp\left(\frac{k\sqrt{v_z^2\sin^2\theta+v_p^2\cos^2\theta}}
{T}\right)}
{\left[1+\exp\left(\frac{k\sqrt{v_z^2\sin^2\theta+v_p^2\cos^2\theta}}
{T}\right)\right]^2}
\right\},\label{Eq_C_v_free}
\end{eqnarray}
\end{widetext}
where the fermion velocities $v_{z, p}$ are fixed constants for the free situation.
In sharp contrast, while taking into account the interplay between fermion-fermion couplings
and disorder scatterings, the fermion velocities $v_{z, p}$ are no longer constants but energy-dependent, namely
coupling with other interaction parameters via the coupled RG equations~(\ref{Eq_v_RG})-(\ref{Eq_Delta_RG}). Under this
circumstance, the specific heat can be rewritten as,
\begin{widetext}
\begin{eqnarray}
\frac{C_V^{\mathrm{int}}(T)}{\Lambda_0^2 k_F}
=
\frac{1}{\pi^2 T^2}
\int_{e^{-l_c}}^{1} \int_0^\pi dk d\theta
\left\{
\frac{k^3[v_z^2(l)\sin^2\theta+v_p^2(l)\cos^2\theta]
\exp\left(\frac{k\sqrt{v_z^2(l)\sin^2\theta+v_p^2(l)\cos^2\theta}}
{T}\right)}
{\left[1+\exp\left(\frac{k\sqrt{v_z^2(l)\sin^2\theta+v_p^2(l)\cos^2\theta}}
{T}\right)\right]^2}
\right\},\label{Eq_C_v_int}
\end{eqnarray}
\end{widetext}
where the fermion velocities $v_{z,p}(l)$ dependent on the energy scale denoted by $l$, and the variable
$l_c$ stands for the critical energy scale.

By combining Eqs.~(\ref{Eq_C_v_free})-(\ref{Eq_C_v_int}) and the RG Eqs.~(\ref{Eq_v_RG})-(\ref{Eq_Delta_RG}),
and paralleling the analogous numerical analysis for DOS and compressibility, we then obtain the temperature-dependent
properties of specific heat in the vicinity of all eight kinds of FPs, which are presented in Fig.~\ref{CV-FP-1-5} and Fig.~\ref{CV-multiply-FPs}.

At first, let us examine the influence of initial values of anisotropy of fermion velocities ($v_{z0}/v_{p0}$) on the
specific heat around distinct kinds of FPs.
Fig.~\ref{CV-FP-1-5}(a) displays the $T$-dependent behavior of $C_V$ around the FP-1 (the free case). It can be
found that the specific heat gradually rises as $T$ increases, exhibiting $C_V(T)\propto T^{2+\delta}$ with $\delta\ll1$.
Although these basic results are robust against the initial anisotropy of fermion velocities,
the specific values of $C_V(T)$ can be considerably suppressed in the presence of anisotropy of fermion velocities, which
are in well agreement with the DOS and compressibility. Paralleling the similar analysis,
we notice that the basic results around FP-2,3,4,6,7,8 are analogous to those around FP-1.
Considering Fig.~\ref{CV-FP-1-5}(b), which illustrates the behavior of $C_V(T)$ in the vicinity of FP-5, its is evident that,
in the isotropic case, $C_V(T)$ increases more rapidly than its counterpart in FP-1.
However, when the fermion velocities are anisotropic, we can find that the specific heat manifestly
deviates from $C_V(T)\propto T^{2+\delta}$ and exhibits an approximately linear dependence on temperature.
This qualitative deviation of $T$-dependent of specific heat nearby FP-5 from
the FP-1, in which the system is under the non-interacting circumstance and the quasiparticles
obey the Fermi-liquid theory, suggests that the quasiparticles as approaching FP-5
are unlikely still under the scope of the Fermi liquid~\cite{Mahan1990Book}.
As a consequence, the behavior of specific heat nearby FP-5 may indicate the emergence of some non-Fermi-liquid behavior
induced by the interplay of fermion-fermion interactions and disorder scatterings.


Next, we select a specific value of $v_{z0}/v_{p0}$ to present a direct comparison of the $T-$dependence of specific heat
around all FPs illustrated in Fig.~\ref{class-FP-1-2} and Fig.~\ref{class-FP-3-8}. Apparently, one can find that the tendencies of $C_V(T)$ characterized by $C_V(T)\propto T^{2+\delta}$ are qualitatively analogous in the vicinity of FP-1, FP-2, FP-3, FP-4, FP-6, FP-7, and FP-8.
In sharp contrast, nearby FP-5, the quadratic temperature dependence is destroyed by the ferocious fluctuations.
Specifically, it approximately exhibits a linear dependence on $T$, which is consistent with Fig.~\ref{CV-FP-1-5}(b).
Besides, the specific heat gains much bigger values around FP-5 compared to the other types of FPs at a certain fixed
temperature. This may be ascribed to the intricate competition between
the fermionic interaction and the disorder scatterings as well as
the thermal fluctuations.



\begin{figure}
\centering
\includegraphics[width=3in]{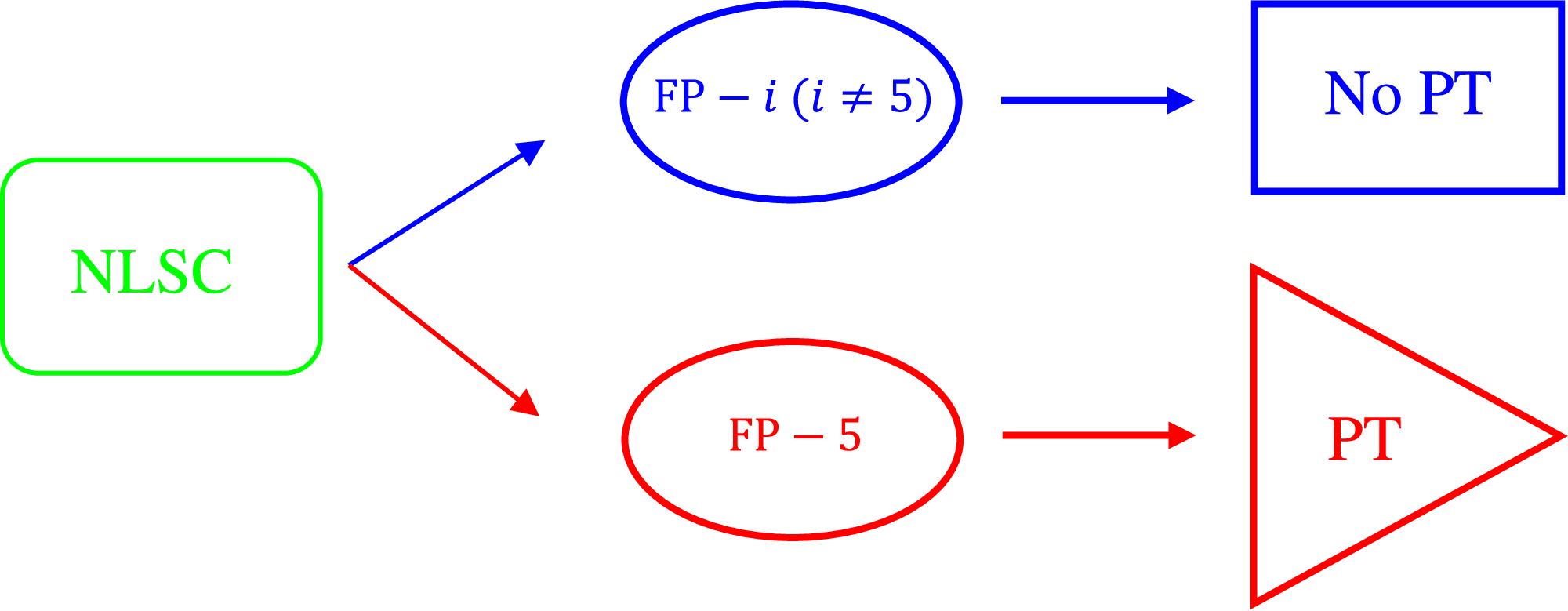}\\
\caption{(Color online) Schematic illustration for the qualitative results of the topological nodal-line superconductor (NLSC)
as approaching the distinct kinds of FPs in Fig.~\ref{class-FP-1-2}
and Fig.~\ref{class-FP-3-8} (here, PT denotes the emergence of certain a phase transition). }\label{sum_all}
\end{figure}

\begin{figure}[htpb]
  \centering
  \includegraphics[width=3.5in]{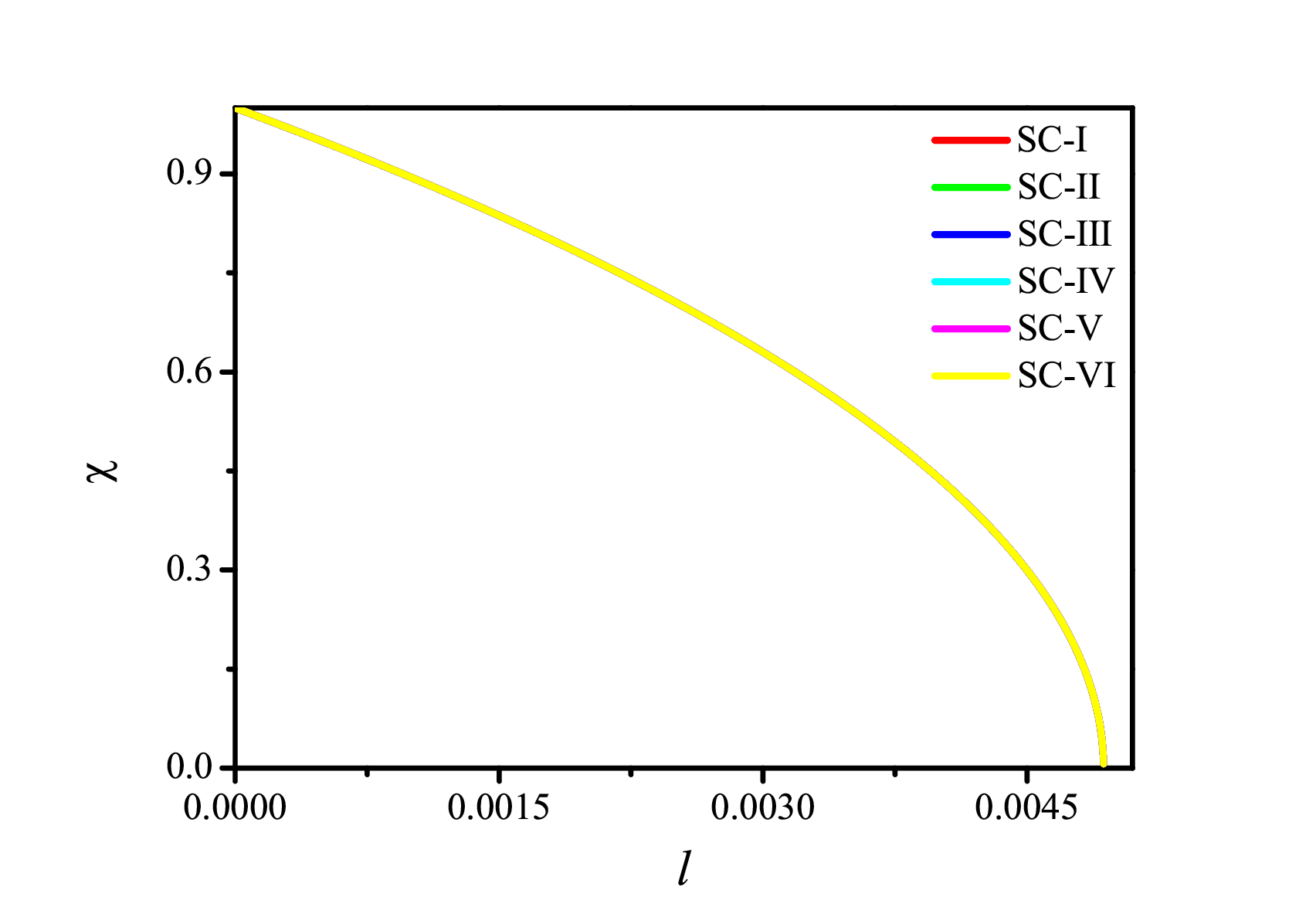}\\
  \caption{(Color online)  Energy-dependent susceptibilities
  of six distinct candidate states as approaching FP-1 for $v_{z0}/v_{p0}=0.5$ and $\theta=\theta'=\pi/3$
  (the basic results for the FP-2, 3, 4, 6, 7, 8 are similar and not shown for simplicity).}
  \label{chi-FP-1}
\end{figure}

\begin{figure*}[htpb]
\centering
\subfigure[]{\includegraphics[width=3in]{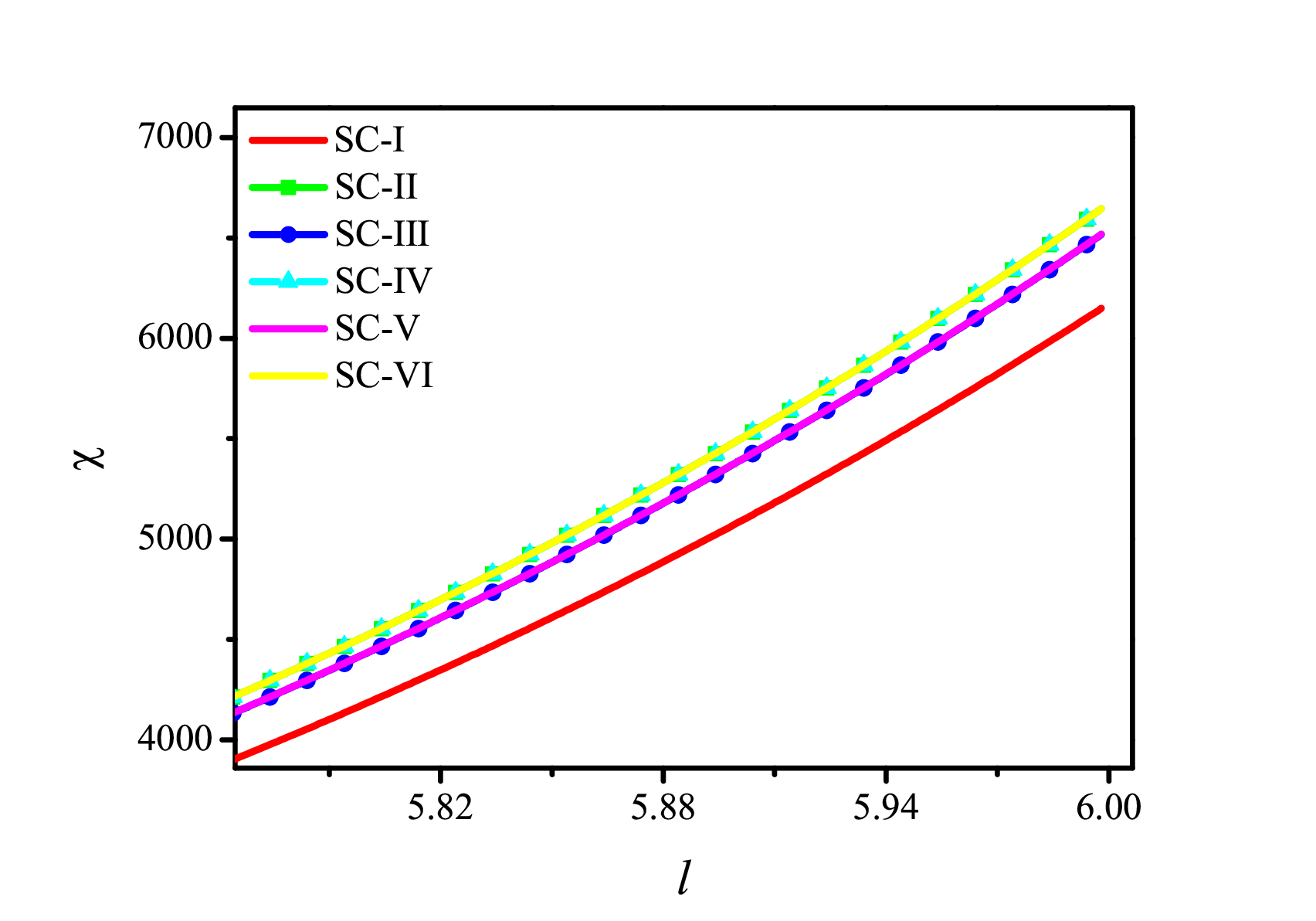}}
\hspace{0.01\linewidth}
\subfigure[]{\includegraphics[width=3in]{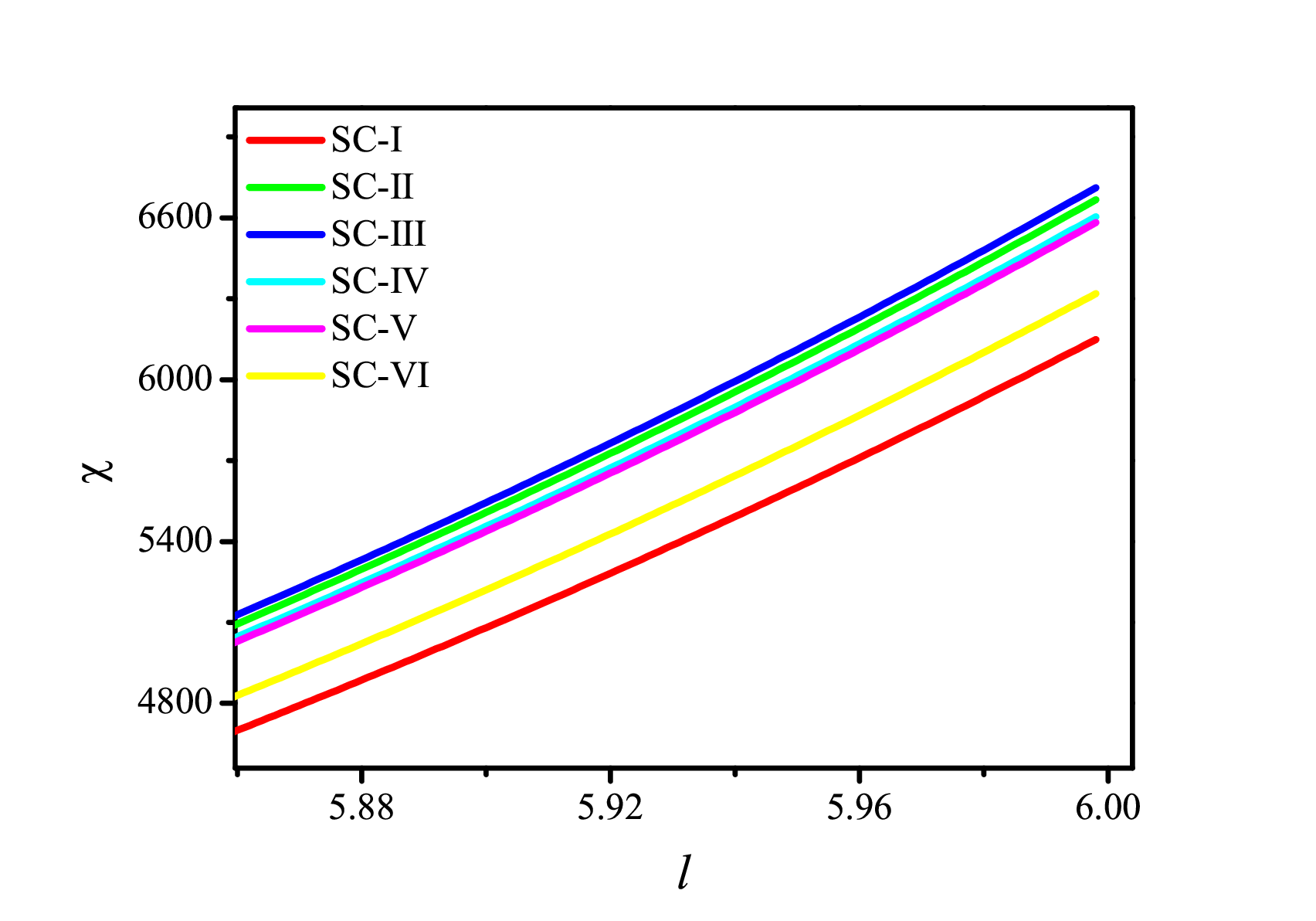}}
\vfill
\subfigure[]{\includegraphics[width=3in]{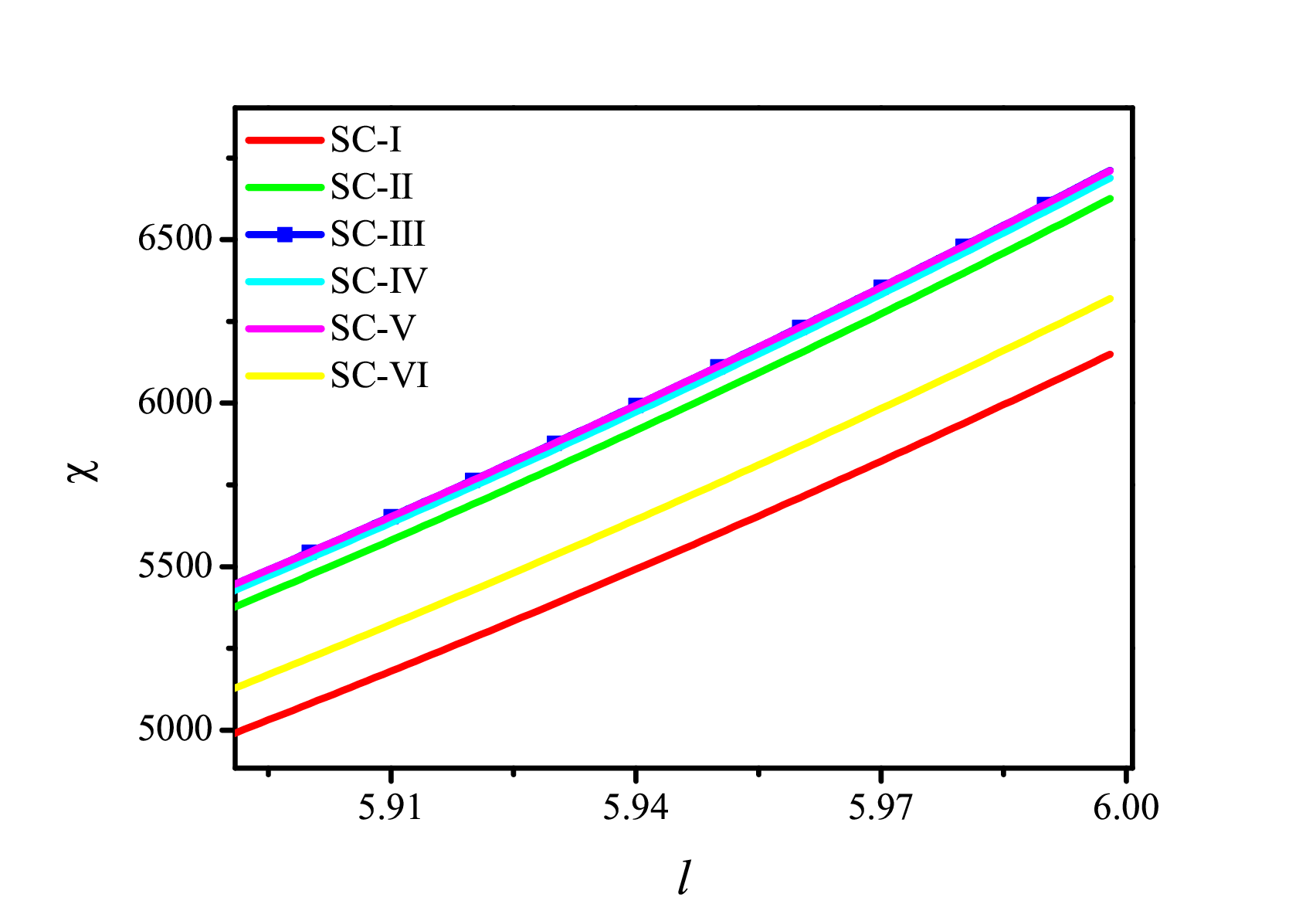}}
\hspace{0.01\linewidth}
\subfigure[]{\includegraphics[width=3in]{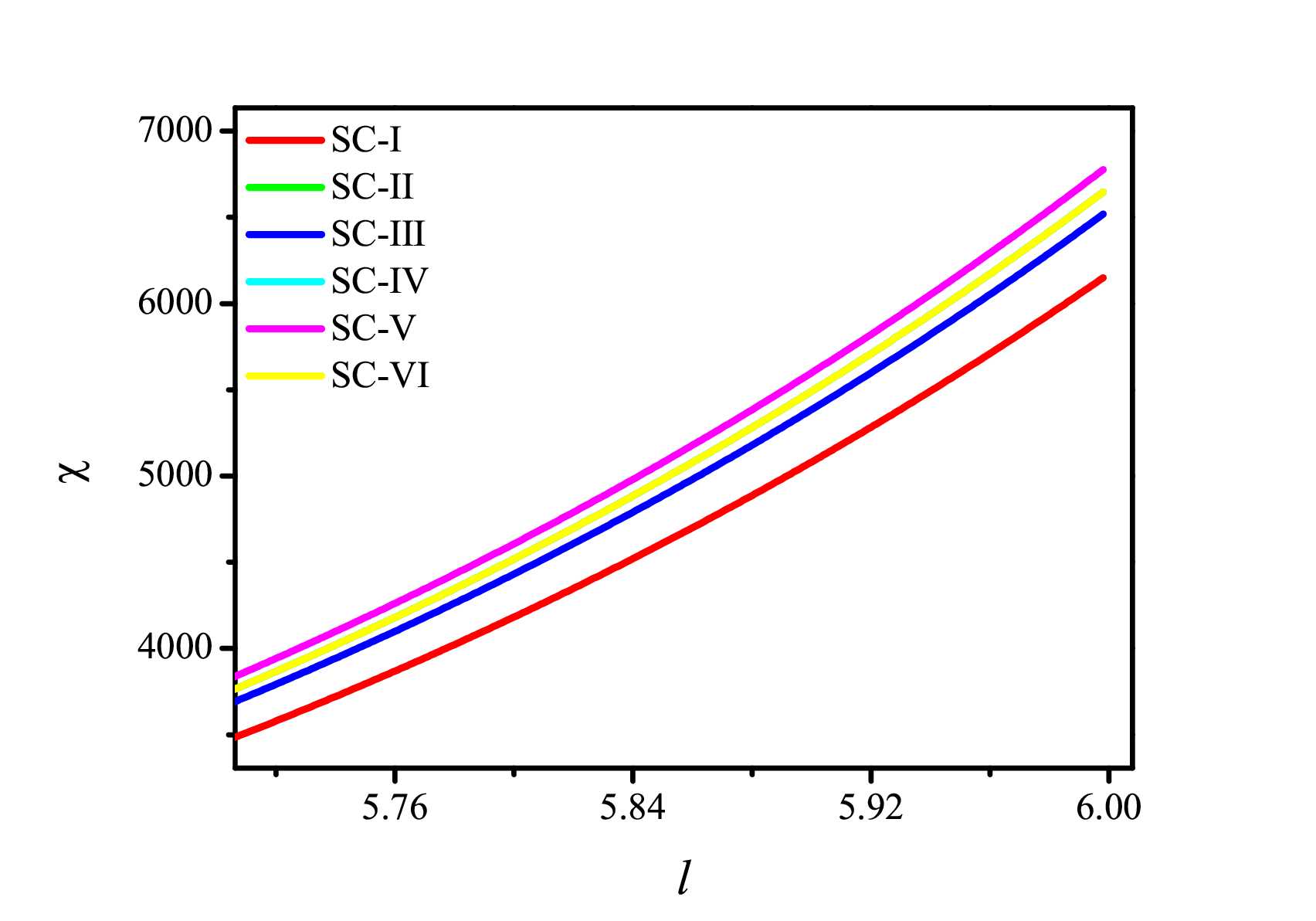}}\\
\caption{(Color online) Energy-dependent susceptibilities
  of six distinct candidate states  as approaching FP-5 for $v_{z0}/v_{p0}=0.5$ and
  $\theta=\pi/3$ with: (a) $\theta'=0$, (b) $\theta'=\pi/3$,
  (c) $\theta'=2\pi/3$, and (d) $\theta'=\pi$, respectively.}
\label{FP-5-vzvp-0.5}
\end{figure*}

Before closing this section,  it is of necessity to give several comments on the following two issues.

At first, we present the reasons behind the peculiar behavior around the FP-5.
As schematically illustrated in Fig.~\ref{class-FP-1-2} and Fig.~\ref{class-FP-3-8},
the fermion-fermion interactions become irrelevant and saturate at a small finite value as approaching FP-2,4,7,8 and FP-6, respectively.
For the FP-3, despite the fermion-fermion interactions are not straightforwardly irrelevant, the increase of disorder strengths suppresses the
impacts of fermion-fermion interactions. In consequence, these render that the physical quantities only acquire quantitative deviations
from their FP-1 (the non-interacting FP) counterparts.In sharp contrast, as shown in Fig.~\ref{class-FP-3-8}(c) for FP-5, the
fermion-fermion couplings increase and meanwhile the strengths of disorder decrease.
Consequently, nearby this kind of FP, the fermion-fermion interactions play a dominant role in shaping the physical implications in the low-energy regime.
As the fermion velocities explicitly enter into the expressions of physical quantities, this difference would also distinguish the
FP-5 from the other FPs.

Subsequently, we move to explain the distinction of properties between the isotropic and
anisotropic cases. In this work, the focus is put on the nodal-line superconductors which are characterized by
an anisotropy of fermion velocities ($v_{z0}/v_{p0}\neq 1$). In comparison, the limited case with the isotropic
ferimonic velocities ($v_z/v_p=1$) is just an ideal model and not applicable for the realistic materials.
For the sake of completeness, we regard the isotropic case as a reference point and also show the related results for such limited case in
Figs.~\ref{DOS-FP-1}-\ref{kappa-FP-8} and Fig.~\ref{CV-FP-1-5},
from which one can figure out the sensitivity of physical behavior as deviating from the isotropic case.
To proceed, we argue that three ingredients may accounting for the distinctions between the isotropic and anisotropic cases.
On one hand, from a physical perspective, an isotropic case ($v_{z0}/v_{p0}=1$) is qualitative different from an anisotropic case ($v_{z0}/v_{p0}\neq1$). Two totally different cases exhibit quite distinct results would be reasonable. On the other, from a mathematical point of view,
the flows of coupled RG equations may be very sensitive to the ratio of fermion velocities, in particular $v_{z0}/v_{p0}=1$ or $v_{z0}/v_{p0}\neq1$.
This accordingly influence the energy-dependent competition between fermion-fermion interaction and disorder scatterings, which then imposes on the
expressions of physical quantities.  Besides, the expressions of physical quantities themselves may be also very susceptible to the ratio
of fermion velocities. All of these together cause the remarked difference between such two distinct cases.

\section{Potential phase transitions nearby fixed points}\label{Sec_phase_transitions}

As presented in the Sec.~\ref{Sec_fixed_points}, there exist eight distinct sorts
of fixed points due to the interplay of fermionic interactions and disorder scatterings.
In particular, some of them are accompanied by the divergences of the interaction
parameters, which generally induce certain instabilities and are linked to the
phase transitions with symmetry breakings~\cite{Vafek2012PRB,Vafek2014PRB,Wang2017QBCP, Maiti2010PRB, Altland2006Book,Vojta2003RPP,Halboth2000RPL,Halboth2000RPB,Eberlein2014PRB,Chubukov2012ARCMP,
Nandkishore2012NP,Chubukov2016PRX,Roy2018PRX,Wang2020NPB,Fu-Wang2023}.
Within this section, we endeavor to determine the dominant instabilities and associated
phase transitions as approaching all kinds of fixed points, which would provide
useful clues to study the low-energy properties of line-nodal superconductors.

\subsection{Candidate states and source terms}

Given the preserved symmetries of the topological nodal-line superconductor (NLSC) and its unique low-energy excitations~(\ref{Eq_H_0-3})
presented in Sec.~\ref{Sec_model}, such kind of system may experience a certain phase transition from the NLSC phase
to another SC phase below the superconducting critical temperature. Table~\ref{table_source} collects the six potential candidate states
due to the time-reversal symmetry breaking~\cite{Moon2017PRB}.

In order to describe these distinct states, we bring out the following source terms
~\cite{Maiti2010PRB,Vafek2014PRB,Chubukov2016PRX},
\begin{eqnarray}
S_{\mathrm{source}} &=& \int d\tau\int d^3\mathbf{x}
\left(\sum^6_{i=1}g_i\Psi^\dagger \mathcal{M}_i
\Psi\right).\label{Eq_S_source}
\end{eqnarray}
Hereby, the matrix $\mathcal{M}_i$ distinguishes the fermion bilinear associated with
the candidate states in Table~\ref{table_source}, and $g_i$ represents the
corresponding order parameter. This source term~(\ref{Eq_S_source}) together with the effective action~(\ref{Eq_S_eff})
gives rise to a renormalized effective action
\begin{eqnarray}
S'_{\mathrm{eff}} = S_{\mathrm{eff}} + S_{\mathrm{source}}.
\label{Eq_S_eff_new}
\end{eqnarray}
Paralleling the similar steps for establishing the RG equations of the fermion-fermion interactions in Sec.~\ref{Sec_RG_equations},
we start with the renormalized effective action~(\ref{Eq_S_eff_new}) to evaluate the one-loop corrections to the source
strength $g_i$, and eventually arrive at the energy dependent equations as follows
\begin{eqnarray}
\frac{dg_i}{dl}=\mathcal{G}(v_{z,p},\lambda_j,\Delta_k,g_i),\label{g_i_exp}
\end{eqnarray}
with fermion-fermion interaction $\lambda_j$ ($j=1-6$) and the disorder strength
$\Delta_k$ ($k=1,2,3,41,42,43$). The detailed calculations and the concrete expressions of $\mathcal{G}_i$
are provided in Appendix~\ref{Sec_appendix-one-loop-source_terms}.


\subsection{Susceptibilities and leading phases}

In order to judge which state in Table~\ref{table_source} is the most favourable one around different kinds of FPs
in Fig.~\ref{class-FP-1-2} and Fig.~\ref{class-FP-3-8}, we are suggested to calculate and compare
the susceptibilities that are associated with each possible order as the NLSC approaches
all kinds of FPs~\cite{Vafek2012PRB,Vafek2014PRB,Wang2017QBCP}.
To this goal, we introduce the expression for the susceptibility as follows~\cite{Vafek2014PRB}
\begin{eqnarray}
\delta\chi=\frac{\partial^2 f}{\partial g(0)\partial g^*(0)},\label{Eq_chi}
\end{eqnarray}
where $f$ denotes the free energy density. With all these
in hand, we are now in a suitable position to compute the susceptibilities of
all candidates states and determine the leading phases nearby the underlying
FPs.


By combining the formula of susceptibility~(\ref{Eq_chi}) and the energy-dependent flow of source strength~(\ref{g_i_exp})
as well as RG equations of interaction parameters~(\ref{Eq_v_RG})-(\ref{Eq_Delta_RG}), we perform the careful numerical analysis
as approaching all FPs and notice that the fates of susceptibilities qualitatively cluster into two distinct scenarios as schematically
presented in Fig.~\ref{sum_all}.

As presented in Sec.~\ref{Sec_fixed_points} and Sec.~\ref{Sec_critical_behavior}, the tendencies of fermion-fermion interactions
are particularly crucial to the low-energy critical behavior.
On one hand, as shown in Fig.~\ref{class-FP-1-2} and Fig.~\ref{class-FP-3-8},
the strengths of fermion-fermion interactions are suppressed by the disorder scatterings nearby FP-3,
saturate at a finite value around FP-6, and are even irrelevant in the vicinity of FP-1,2,4,7,8, respectively.
On the other hand, considering the FP-5, the fermionic couplings win against the disorder scattering and
play a dominant role in pinning down the low-energy physics.

According to the numerical analysis, we notice that, for the former,
the susceptibilities of all potential phases in Table~\ref{table_source} gradually decrease
and vanish with lowering the energy scales. As the basic results are analogous, we take FP-1 for an instance and
present the energy-dependent evolutions of susceptibilities in Fig.~\ref{chi-FP-1}.
It signals that there does not exists any instability around these FPs, which implies the current state
is robust enough. In consequence, there does not exist any
phase transition below the critical temperature of the NLSC.
In sharp contrast, when approaching the latter, we notice that
certain instabilities are triggered by the divergent fermion-fermion interactions. The emergence of
instabilities consequently induce some phase transition from
the NLSC to one of candidate states provided in Table~\ref{table_source}, which we denote the leading phase.
To proceed, let us put our focus on such nontrivial scenario.

\begin{figure*}[htpb]
\centering
\subfigure[]{\includegraphics[width=3in]{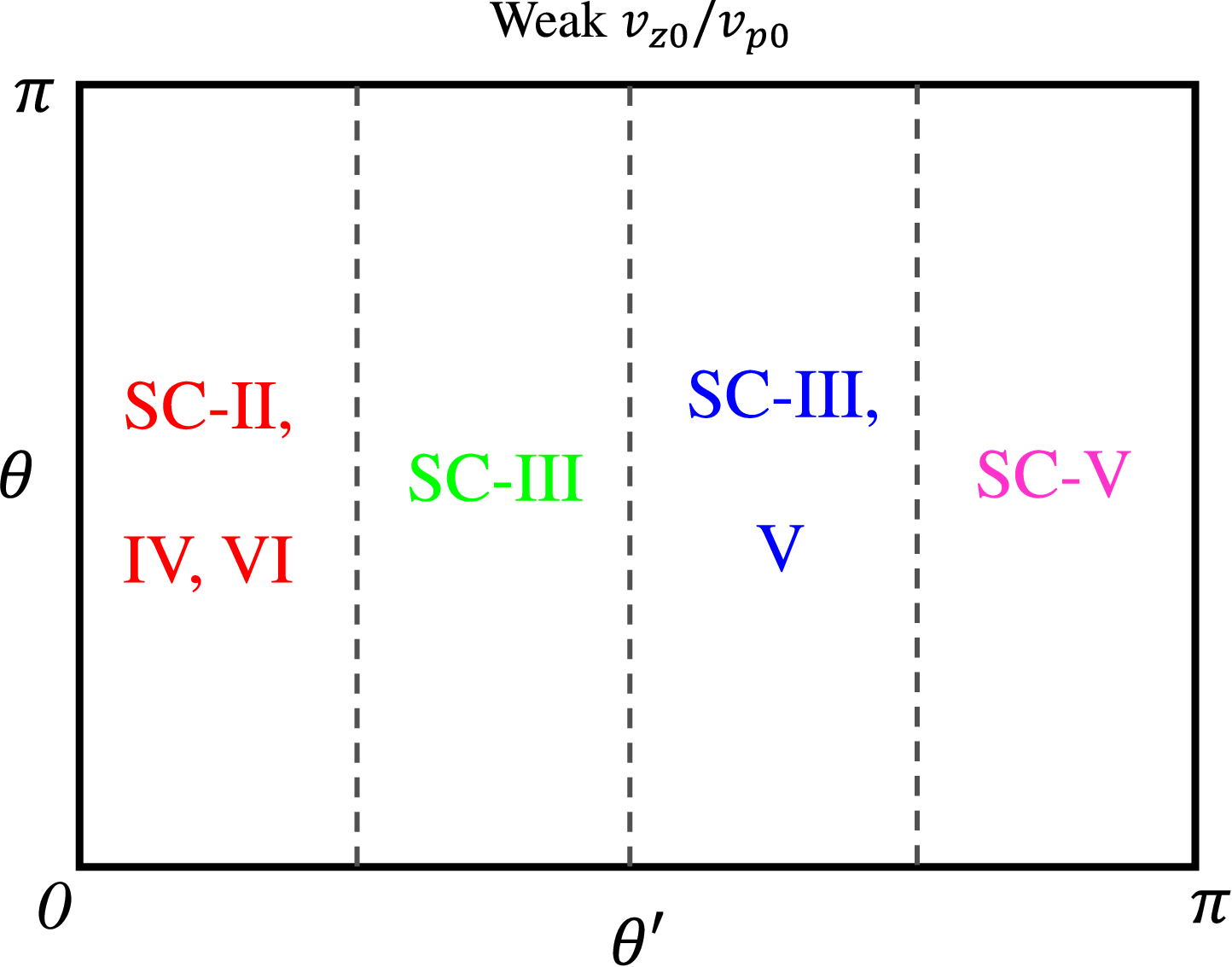}}
\hspace{0.001\linewidth}
\subfigure[]{\includegraphics[width=3in]{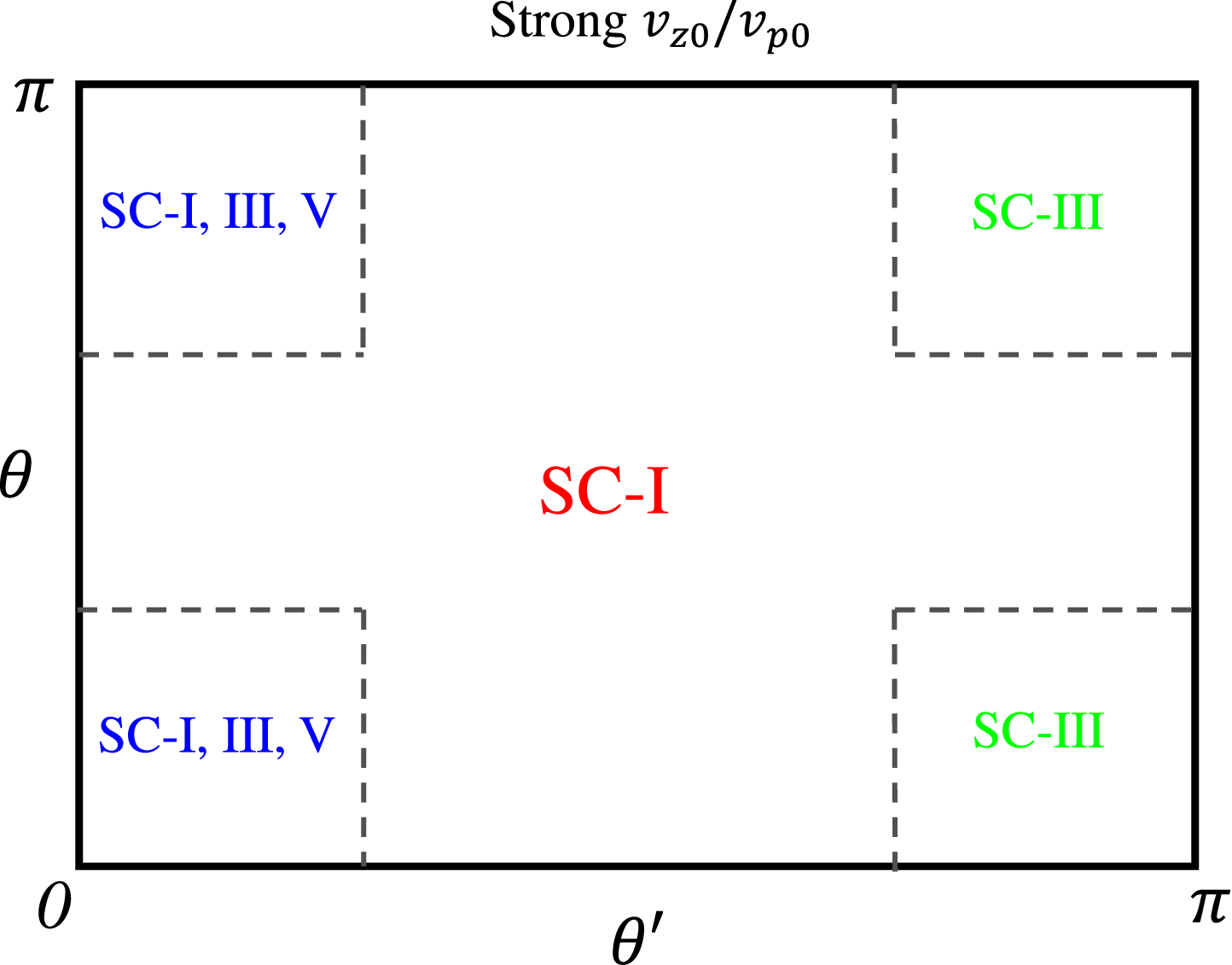}}\\
\caption{(Color online) Schematic illustration of the leading phases of the phase transition
as depicted in Fig.~\ref{sum_all} in the vicinity of FP-5 with variations of $\theta'$ and $\theta$ for: (a) a weak anisotropy of $v_{z0}/v_{p0}$ and
(b) a strong anisotropy of $v_{z0}/v_{p0}$, respectively.}\label{sum_FP-5}
\end{figure*}

Since the anisotropy of fermion velocities is one of unique features for the nodal-line superconductors, one can expect
the tendencies of instabilities around FP-5 would be closely pertinent to the ratio of $v_{z0}/v_{p0}$. The numerical
analysis firmly corroborates this point.
According to the numerical results, we realize that the leading phases are primarily dependent upon the initial anisotropy of
fermion velocities and the competition between two angles ($\theta$ and $\theta'$, which specify the momentum directions of
nodal-line fermions involved in the fermion-fermion interactions).
At first, let us consider a weak anisotropy of fermion velocities, for instance $v_{z0}/v_{p0}=0.5$.
Fig.~\ref{FP-5-vzvp-0.5} with a representative value of $\theta=\pi/3$ shows that the dominant phases around the FP-5 are
sensitive to the value of $\theta'$. With variation of $\theta'$, the leading phase can either be a single state (SC-III or SC-V)
or a mixed state, namely a two-degenerate state SC-(III,V) or three-degenerate sate SC-(II, IV, VI). We have checked that these
basic results are insusceptible to $\theta$. Consequently, the $\theta'$ dominates the competition with $\theta$ and
play a crucial role in determining the leading phases as schematically summarized in Fig.~\ref{sum_FP-5}(a).
Next, we move to the situation with a strong anisotropy of fermion velocities (such as $v_{z0}/v_{p0}\leq0.1$).
In this circumstance, the competition between $\theta$ and $\theta'$ is more intricate and thus renders that the
leading phases heavily hinge upon both $\theta$ and $\theta'$ as schematically presented in Fig.~\ref{sum_FP-5}(b).
In other words, armed with the strong anisotropy of fermion velocities,
the directions of the fermionic momenta impose a significant restriction on
the selection of dominant phases.
When $\theta$ is adjacent to $0$ or $\pi$, the three-degenerate state SC-(I, III, V) and
SC-III become dominant at a small and big $\theta'$, respectively. Besides, the leading phase
is the SC-I state around FP-5 for other regions. As a result, one can figure out that both
the anisotropy of the fermion velocities and the directions of fermion momenta
can bring a qualitative impact on the leading state of the phase transition around FP-5.
In principle, the nontrivial critical behavior would be generally associated with the underlining instability (phase transition) at which the
quantum fluctuations are of particular importance. In this sense,
the existence of a potential phases transition nearby FP-5 is also in well consistent with the distinguished critical
behavior around FP-5 compared to that of
other FPs presented in Sec.~\ref{Sec_critical_behavior}.

To be brief, as the NLSC approaches FP-1,2,3,4,6,7,8, the fermion-fermion interactions are either irrelevant or suppressed by
the disorder scatterings, and henceforth there does not emerge any nontrivial instability as displayed in Fig.~\ref{sum_all}.
However, the fermionic couplings dominant over the disorder scatterings and drive certain instabilities as approaching FP-5.
This consequently suggests that the NLSC can undergo a phase transition to a candidate state
listed in Table~\ref{table_source}. For simplicity,
Fig.~\ref{sum_FP-5} schematically illustrates the leading phases induced by the potential phase
transition nearby FP-5.

\section{Summary}\label{Sec_summary}

In summary, this work delves into the intricate consequences
of the interplay between short-range fermion-fermion interactions and disorder scatterings on the critical
physical behavior in the low-energy regime beneath the superconducting dome of nodal-line superconductors.
In order to treat all degrees of freedom on the same footing, we adopt the powerful momentum-shell RG
approach~\cite{Wilson1975RMP,Polchinski9210046,Shankar1994RMP} to establish the coupled energy-dependent RG equations via collecting
all one-loop corrections arising from potential interactions. After performing the numerical analysis of these equations, we identify the
fixed points of interaction parameters in the low-energy regime, and address a number of unique behavior of physical implications
as well as present instabilities accompanied by certain phase transitions as approaching all potential fixed points.

At the outset, we analyze the coupled RG flow equations~(\ref{Eq_v_RG})-(\ref{Eq_Delta_RG}) and find that the
system evolves towards eight distinct sorts of fixed points as schematically shown in Fig.~\ref{class-FP-1-2} and Fig.~\ref{class-FP-3-8}
in the low-energy regime. These fixed points capture the essential physical information that encapsulates the intimate competition between
fermion-fermion interactions and disorder scatterings, and henceforth
dictate the low-energy critical behavior of our nodal-line superconductors.

To proceed, we thoroughly investigate the low-energy fates of the DOS and compressibility as well as specific heat when the system
approaches these fixed points. As to the DOS, nearby the FP-1 for the non-interacting case, it exhibits a linear dependence on frequency
and preserves the complete symmetry between $\omega>0$ and $\omega<0$ as depicted in Fig.~\ref{DOS-FP-1}.
In comparison, as we approach the other kinds of FPs in the presence of fermion-fermion interactions and disorder scatterings,
the DOS gradually increases at first and then decreases with tuning up the frequency as shown in Fig.~\ref{DOS-FP-3}-Fig.~\ref{DOS-FP-5}.
Besides, the contributions from all interaction parameters gives rise to a non-zero finite DOS at $\omega=0$,
and the initial anisotropy of fermion velocities are able to quantitatively impact the tendency of DOS.
After paralleling the analogous analysis of DOS, we realize that the chemical potential ($\mu$) dependence
of compressibility ($\kappa$) exhibits similar tendencies to its temperature-dependent DOS counterpart.
To be concrete, Fig.~\ref{kappa-FP-1} shows $\kappa(\mu)\propto\mu$ around FP-1 , but instead
Fig.~\ref{kappa-FP-2}-Fig.~\ref{kappa-FP-1-8} present that $\kappa$ goes up below a critical value $\mu_c$ and then
decreases at $\mu>\mu_c$ with tuning up $\mu$ near other FPs. Next, with respect to the specific heat of quasiparticles, we notice that
the tendencies of $C_V(T)$ depicted in Fig.~\ref{CV-FP-1-5}(a) and Fig.~\ref{CV-multiply-FPs}
are followed by $C_V(T)\propto T^{2+\delta}$ as approaching the FP-$i$ with $i\neq5$ and $\delta\ll1$.
In sharp contrast, nearby FP-5, the quadratic temperature dependence is sabotaged by the ferocious fluctuations
and then replaced by an approximately linear dependence on $T$ as clearly illustrated in Fig.~\ref{CV-FP-1-5}(b),
which signals the emergence of non-Fermi-liquid behavior. In addition, the related specific heat $C_V(T)$ of this case
is much bigger compared to those of the other FPs at a certain fixed
temperature.

Moreover, we examine the potential instabilities and phase transitions around all the FPs.
To this end, we introduce the source terms of possible candidate states and then calculate all the one-loop corrections
to construct their energy-dependent equations. After combining these flow equations and the RG equations of interaction parameters,
we obtain the energy-dependent susceptibilities of the potential candidate states. As corollary, this suggests a certain phase transition around FP-5
from the nodal-line superconducting state to another superconducting state below the critical temperature of nodal-line superconductor
as illustrated in Fig.~\ref{sum_FP-5}. In particular, the specific state accompanied by such a phase transition is primarily determined by the
intricate combination of two momentum orientations of quasiparticles participated in fermion-fermion interactions. To recapitulate,
we hope these results would improve our understandings of the 3D nodal-line superconductors and
provide helpful clues for further experiments that are associated with these critical behavior.

\section*{ACKNOWLEDGEMENTS}

We thank Yi-Sheng Fu and Wen Liu for the helpful discussions.
J.W. was partially supported by the National Natural
Science Foundation of China under Grant No. 11504360.

\section*{Data Availability Statement}

Data Availability Statement: No Data associated in
the manuscript.

\appendix

\section{One-loop RG equations}\label{Sec_appendix-one-loop-RGEqs}

\begin{figure*}
\centering
\includegraphics[width=6.15in]{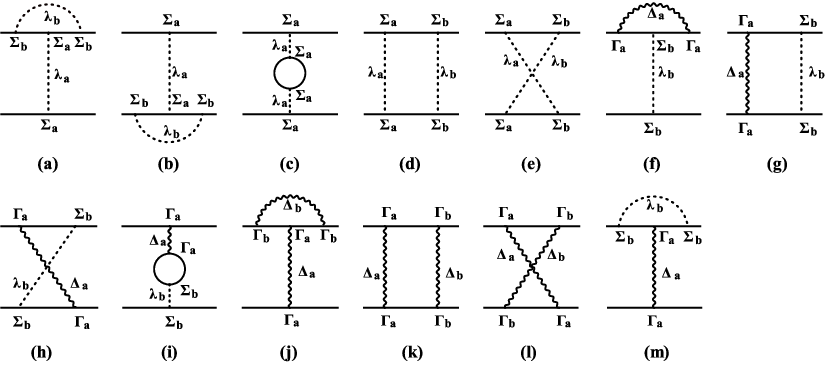}
\vspace{-0.2cm}
\caption{One-loop corrections to the fermion-fermion couplings (a)-(i) and
the fermion-disorder strengths (j)-(m) (the solid, dashed
and wavy lines represent the fermion propagator, fermion-fermion interactions
and disorder scattering, respectively)~\cite{Wang2020PRB,Roy-Saram2016PRB,BCW2023EPJP}.}\label{Fig_one-loop-ff-inter}
\end{figure*}

\begin{figure}
\centering
\includegraphics[width=3.25in]{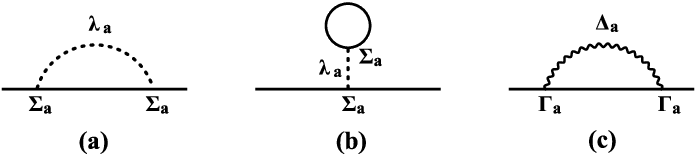}
\vspace{-0.2cm}
\caption{One-loop corrections to the fermion propagator
due to the fermion-fermion interactions (a) and (b)
as well as disorder scatterings (c) (the solid, dashed,
and wavy lines represent the fermion propagator, fermion-fermion interaction,
and disorder scattering, respectively).}\label{Fig_one-loop-self-energy}
\end{figure}

Following the spirit of RG approach~\cite{Wilson1975RMP,Polchinski9210046,Shankar1994RMP}
and employing the RG rescaling transformations~(\ref{Eq_RG-scaling-1})-(\ref{Eq_RG-scaling-2}),
and taking into account all the one-loop corrections as depicted in Fig.~\ref{Fig_one-loop-ff-inter}
and Fig.~\ref{Fig_one-loop-self-energy}, which contain the interplay between fermion-fermion interactions and disorder scatterings
(cf. our previous work~\cite{BCW2023EPJP} for details), we finally derive
the coupled RG flow equations of all coupling parameters,
\begin{widetext}
\begin{small}
\begin{eqnarray}
\frac{dv_z}{dl}
&=&[4\mathcal{C}_{1}\lambda_{2}-\mathcal{C}_{2}
(\Delta_{1}+\Delta_{2}+\Delta_{3}
+\Delta_{41}+\Delta_{42}+\Delta_{43})]v_z,\label{Eq_RG_v_z}\\
\frac{dv_p}{dl}
&=&[4\mathcal{C}_{1}\lambda_{1}-\mathcal{C}_{2}
(\Delta_{1}+\Delta_{2}+\Delta_{3}
+\Delta_{41}+\Delta_{42}+\Delta_{43})]v_p,\label{Eq_RG_v_perp}\\
\frac{d\lambda_1}{dl}
&=&2\lambda_1\Bigl[-\frac{1}{2}+\mathcal{C}_{3}
\left(
-3\lambda_{1}-2\lambda_{2}-\lambda_{3}+\lambda_{4}
+\lambda_{5}-\lambda_{6}
\right)+
\mathcal{C}_{7}(\Delta_{1}
-\Delta_{2}
-7\Delta_{3}
-\Delta_{41}
-\Delta_{42}
-\Delta_{43})\nonumber\\
&&-\mathcal{C}_{2}
(\Delta_{1}+\Delta_{2}+\Delta_{3}
+\Delta_{41}+\Delta_{42}+\Delta_{43})\Bigr],\\
\frac{d\lambda_2}{dl}
&=&2\lambda_2\Bigl[-\frac{1}{2}+\mathcal{C}_{4}
\left(
-2\lambda_{1}-3\lambda_{2}+\lambda_{3}+\lambda_{4}
-\lambda_{5}-\lambda_{6}
\right)+
\mathcal{C}_{7}(-\Delta_{1}
+\Delta_{2}
+\Delta_{3}
-\Delta_{41}
-\Delta_{42}
-\Delta_{43}
)\nonumber\\
&&-\mathcal{C}_{2}
(\Delta_{1}+\Delta_{2}+\Delta_{3}
+\Delta_{41}+\Delta_{42}+\Delta_{43})\Bigr],\\
\frac{d\lambda_3}{dl}
&=&2\lambda_3\Bigl[-\frac{1}{2}+\mathcal{C}_{4}
\left(
-2\lambda_{1}+\lambda_{2}-3\lambda_{3}-\lambda_{4}
+\lambda_{5}+\lambda_{6}
\right)
+
\mathcal{C}_7(-\Delta_1
+\Delta_2
+\Delta_3
+\Delta_{41}
+7\Delta_{42}
+\Delta_{43})\nonumber\\
&&
-\mathcal{C}_{2}
(\Delta_{1}+\Delta_{2}+\Delta_{3}
+\Delta_{41}+\Delta_{42}+\Delta_{43})\Bigr]
-4\mathcal{C}_6\lambda_5\Delta_{41},\\
\frac{d\lambda_4}{dl}
&=&
\lambda_4[-1-4\mathcal{C}_{2}
(\Delta_{41}+\Delta_{42})]
-4\mathcal{C}_5(\lambda_5\Delta_2
+\lambda_6\Delta_3)
-2\lambda_{6}(\mathcal{C}_{4}\lambda_{2}
+\mathcal{C}_{3}\lambda_{1}),\\
\frac{d\lambda_5}{dl}
&=&2\lambda_5
\Bigl[-\frac{1}{2}+
\mathcal{C}_3
(\lambda_1-2\lambda_2+\lambda_3
+\lambda_4-3\lambda_5-\lambda_6
)
+
\mathcal{C}_7(\Delta_1
-\Delta_2
+\Delta_3
+\Delta_{41}
+\Delta_{42}
-\Delta_{43}
)\nonumber\\
&&
-\mathcal{C}_{2}
(\Delta_{1}+\Delta_{2}+\Delta_{3}
+\Delta_{41}+\Delta_{42}+\Delta_{43})\Bigr]
+2\mathcal{C}_1\lambda_2\lambda_6-
4(\mathcal{C}_6\lambda_3\Delta_{41}
+\mathcal{C}_5\lambda_4\Delta_2
+\mathcal{C}_5\lambda_6\Delta_1),\\
\frac{d\lambda_6}{dl}
&=&2\lambda_6\Bigl[-\frac{1}{2}+
\mathcal{C}_1
(-\lambda_1-\lambda_2+\lambda_3
+\lambda_4-\lambda_5-3\lambda_6)
-(\mathcal{C}_3\lambda_2+\mathcal{C}_4\lambda_1)
-2\mathcal{C}_2(\Delta_1
+\Delta_2
+\Delta_{41}
+\Delta_{42}
)\Bigr]\nonumber\\
&&+2\mathcal{C}_1\lambda_2\lambda_5
-4\mathcal{C}_5(\lambda_4\Delta_3
+\lambda_5\Delta_1),\label{Eq_RG_lambda_6}\\
\frac{d\Delta_1}{dl}
&=&
2\mathcal{C}_2\Delta_1
(\Delta_1+\Delta_2+\Delta_3
+\Delta_{41}+\Delta_{42}+\Delta_{43})
+8\mathcal{C}_5\Delta_2\Delta_3,\\
\frac{d\Delta_2}{dl}
&=&\Delta_2\Big[2\mathcal{C}_{2}
(-3\Delta_{1}-3\Delta_{2}+\Delta_{3}
+\Delta_{41}+\Delta_{42}+\Delta_{43})+
\mathcal{C}_1
(\lambda_1+\lambda_2+\lambda_3
-\lambda_4+\lambda_5-\lambda_6)
\Bigr]+8\mathcal{C}_5\Delta_1\Delta_3,\label{Eq_RG_Delta-2}\\
\frac{d\Delta_{3}}{dl}
&=&\Delta_{3}\Big[
4\mathcal{C}_7
(\Delta_1-\Delta_2+\Delta_3
-\Delta_{41}-\Delta_{42}-\Delta_{43})
-2\mathcal{C}_{2}
(\Delta_{1}+\Delta_{2}+\Delta_{3}
+\Delta_{41}+\Delta_{42}+\Delta_{43})\nonumber\\
&&+
\mathcal{C}_3
(-\lambda_1+\lambda_2+\lambda_3
-\lambda_4-\lambda_5+\lambda_6)
\Bigr]+8\mathcal{C}_5\Delta_1\Delta_2,\\
\frac{d\Delta_{41}}{dl}
&=&\Delta_{41}\Big[
4\mathcal{C}_7
(-\Delta_1+\Delta_2+\Delta_3
-\Delta_{41}+\Delta_{42}+\Delta_{43})
-2\mathcal{C}_{2}
(\Delta_{1}+\Delta_{2}+\Delta_{3}
+\Delta_{41}+\Delta_{42}+\Delta_{43})\nonumber\\
&&+
\mathcal{C}_4
(\lambda_1-\lambda_2+\lambda_3
+\lambda_4-\lambda_5-\lambda_6)
\Bigr]+8\mathcal{C}_5\Delta_{42}\Delta_{43},\\
\frac{d\Delta_{42}}{dl}
&=&\Delta_{42}\Big[
4\mathcal{C}_7
(-\Delta_1+\Delta_2+\Delta_3
+\Delta_{41}-\Delta_{42}+\Delta_{43})
-2\mathcal{C}_{2}
(\Delta_{1}+\Delta_{2}+\Delta_{3}
+\Delta_{41}+\Delta_{42}+\Delta_{43})\nonumber\\
&&+
\mathcal{C}_4
(\lambda_1-\lambda_2-\lambda_3
+\lambda_4-\lambda_5-\lambda_6)
\Bigr]+
8\mathcal{C}_5\Delta_{41}\Delta_{43},\\
\frac{d\Delta_{43}}{dl}
&=&\Delta_{43}\Big[
4\mathcal{C}_7
(-\Delta_1+\Delta_2+\Delta_3
+\Delta_{41}+\Delta_{42}-\Delta_{43})
-2\mathcal{C}_{2}
(\Delta_{1}+\Delta_{2}+\Delta_{3}
+\Delta_{41}+\Delta_{42}+\Delta_{43})\nonumber\\
&&+
\mathcal{C}_4
(\lambda_1-\lambda_2+\lambda_3
-\lambda_4+\lambda_5+\lambda_6)
\Bigr]+
8\mathcal{C}_5\Delta_{41}\Delta_{42}.\label{Eq_RG_Delta}
\end{eqnarray}
\end{small}
\end{widetext}
Hereby the coefficients $\mathcal{F}$, $\mathcal{J}$, and $\mathcal{K}$ appearing in
equations~(\ref{Eq_v_RG})-(\ref{Eq_Delta_RG}) are explicitly provided in the right hand of above RG equations.
The coefficients $\mathcal{C}_i$ with $i=1-7$ are nominated as follows
\begin{eqnarray}
\mathcal{C}_{1}
&\equiv&\frac{1}{(2\pi)^3}
\int_{0}^{\pi}d\theta
\frac{\pi}
{(\upsilon_{z}^2\sin^2\theta+\upsilon_{p}^2\cos^2\theta)^{1/2}},\label{Eq_coeff-C1}\\
\mathcal{C}_{2}
&\equiv&\frac{1}{(2\pi)^3}
\int_{0}^{\pi}d\theta
\frac{2\pi}
{\upsilon_{z}^2\sin^2\theta+\upsilon_{p}^2\cos^2\theta},\\
\mathcal{C}_{3}
&\equiv&\frac{1}{(2\pi)^3}
\int_{0}^{\pi}d\theta
\frac{\pi\upsilon_{z}^2\sin^2\theta}
{(\upsilon_{z}^2\sin^2\theta+\upsilon_{p}^2\cos^2\theta)^{3/2}},\\
\mathcal{C}_{4}
&\equiv&\frac{1}{(2\pi)^3}
\int_{0}^{\pi}d\theta
\frac{\pi\upsilon_{p}^2\cos^2\theta}
{(\upsilon_{z}^2\sin^2\theta+\upsilon_{p}^2\cos^2\theta)^{3/2}},\\
\mathcal{C}_{5}
&\equiv&\frac{1}{(2\pi)^3}
\int_{0}^{\pi}d\theta
\frac{4\pi\upsilon_{z}^2\sin^2\theta}
{(\upsilon_{z}^2\sin^2\theta+\upsilon_{p}^2\cos^2\theta)^2},\\
\mathcal{C}_{6}
&\equiv&\frac{1}{(2\pi)^3}
\int_{0}^{\pi}d\theta
\frac{4\pi\upsilon_{p}^2\cos^2\theta}
{(\upsilon_{z}^2\sin^2\theta+\upsilon_{p}^2\cos^2\theta)^2},\\
\mathcal{C}_{7}
&\equiv&\frac{1}{(2\pi)^3}
\int_{0}^{\pi}d\theta
\frac{2\pi(\upsilon_{p}^2\cos^2\theta-\upsilon_{z}^2\sin^2\theta)}
{(\upsilon_{z}^2\sin^2\theta+\upsilon_{p}^2\cos^2\theta)^2}.\label{Eq_coeff-C7}
\end{eqnarray}

\section{One-loop equations of source terms}\label{Sec_appendix-one-loop-source_terms}

\begin{figure}[htpb]
\centering
\includegraphics[width=3.1in]{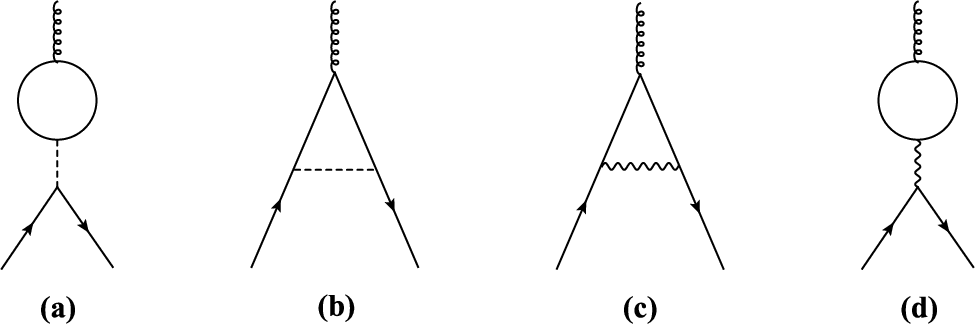}
\vspace{-0.1cm}
\caption{One-loop corrections to the strength of
source terms due to fermion-fermion interactions (a)-(b) and fermion-disorder interaction (c)-(d).
Hereby, the dashed line and wavy line denote the fermion-fermion interactions and fermion-disorder interactions,
respectively.}\label{fig_source-corrections}
\end{figure}

Paralleling the analogous procedures in Appendix~\ref{Sec_appendix-one-loop-RGEqs}, we
calculate the one-loop corrections to the strengths of source terms
as shown in Fig.~\ref{fig_source-corrections}, and then adopt the RG rescalings~(\ref{Eq_RG-scaling-1})-(\ref{Eq_RG-scaling-2})
to finally derive the following flow equations for $g_i$ with $i=1-6$,
\begin{widetext}
\begin{small}
\begin{eqnarray}
\frac{dg_1}{dl}
&=&\Bigl[(1-\eta)+\frac{1}{32\pi^3}[(\lambda_1-\lambda_2-\lambda_3-\lambda_4+\lambda_5+\lambda_6
-\Delta_1-\Delta_2+\Delta_3\nonumber\\
&&-\Delta_{41}-\Delta_{42}-\Delta_{43})I_2
+8(\lambda_2+\Delta_2)I_2+16(\lambda_1+\Delta_3) I_1]\Bigl]g_1,\label{g_1}\\
\frac{dg_2}{dl}
&=&\Bigl[(1-\eta)+\frac{\mathcal{F}_{1}(\theta,\theta')}{4\pi^2}
[2(\lambda_1+\Delta_3)I_1
+(\lambda_2+\Delta_2)I_2]\nonumber\\
&&+\frac{\mathcal{F}_{2}(\theta,\theta')}{4\pi^2}
(\lambda_1-\lambda_2-\lambda_3-\lambda_4+\lambda_5+\lambda_6
-\Delta_1-\Delta_2+\Delta_3-\Delta_{41}-\Delta_{42}-\Delta_{43})I_2\Bigl]g_2,\\
\frac{dg_3}{dl}
&=&\Bigl[(1-\eta)+\frac{\mathcal{F}_{3}(\theta,\theta')}{4\pi^2}
[2(\lambda_1+\Delta_3)I_1
+(\lambda_2+\Delta_2)I_2]\nonumber\\
&&
+
\frac{\mathcal{F}_{4}(\theta,\theta')}{4\pi^2}
(\lambda_1-\lambda_2-\lambda_3-\lambda_4+\lambda_5+\lambda_6
-\Delta_1-\Delta_2+\Delta_3-\Delta_{41}-\Delta_{42}-\Delta_{43})I_2\Bigl]g_3,\\
\frac{dg_4}{dl}
&=&\Bigl[(1-\eta)+\frac{\mathcal{F}_{5}(\theta,\theta')}{4\pi^2}
[2(\lambda_1+\Delta_3)I_1
+(\lambda_2+\Delta_2)I_2]\nonumber\\
&&
+
\frac{\mathcal{F}_{6}(\theta,\theta')}{4\pi^2}
(\lambda_1-\lambda_2-\lambda_3-\lambda_4+\lambda_5+\lambda_6
-\Delta_1-\Delta_2+\Delta_3-\Delta_{41}-\Delta_{42}-\Delta_{43})I_2\Bigl]g_4,\\
\frac{dg_5}{dl}
&=&
\Bigl[(1-\eta)+\frac{\mathcal{F}_{7}(\theta,\theta')}{4\pi^2}
[2(\lambda_1+\Delta_3)I_1
+(\lambda_2+\Delta_2)I_2]\nonumber\\
&&
+
\frac{\mathcal{F}_{8}(\theta,\theta')}{4\pi^2}
(\lambda_1-\lambda_2-\lambda_3-\lambda_4+\lambda_5+\lambda_6
-\Delta_1-\Delta_2+\Delta_3-\Delta_{41}-\Delta_{42}-\Delta_{43})I_2\Bigl]g_5,\\
\frac{dg_6}{dl}
&=&\Bigl[(1-\eta)+\frac{\mathcal{F}_{9}(\theta,\theta')}{4\pi^2}
[2(\lambda_1+\Delta_3)I_1
+(\lambda_2+\Delta_2)I_2]\nonumber\\
&&
+
\frac{\mathcal{F}_{10}(\theta,\theta')}{4\pi^2}
(\lambda_1-\lambda_2-\lambda_3-\lambda_4+\lambda_5+\lambda_6
-\Delta_1-\Delta_2+\Delta_3-\Delta_{41}-\Delta_{42}-\Delta_{43})I_2\Bigl]g_6.\label{g_6}
\end{eqnarray}
\end{small}
\end{widetext}
\begin{small}
\begin{eqnarray}
I_1&=&\int_0^\pi d\theta\frac{v_z v_p\sin\theta
\cos\theta}{(v_z^2\sin^2\theta+v_p^2\cos^2\theta)^{3/2}},\\
I_2&=&\int_0^\pi d\theta\frac{(v_z^2\sin^2\theta-v_p^2\cos^2\theta)}{(v_z^2\sin^2\theta+v_p^2\cos^2\theta)^{3/2}},
\end{eqnarray}
\end{small}
and
\begin{small}
\begin{eqnarray}
\mathcal{F}_{1}(\theta,\theta')
&\equiv&\frac{\sin(4\theta^\prime)\sin(4\theta)}{16\pi},\\
\mathcal{F}_{2}(\theta,\theta')
&\equiv&\frac{\sin(4\theta^\prime)\sin^2(4\theta)}{128\pi},\\
\mathcal{F}_{3}(\theta,\theta')
&\equiv&\frac{\cos(2\theta^\prime)
\cos(2\theta)}{\pi},\\
\mathcal{F}_{4}(\theta,\theta')
&\equiv&\frac{\cos(2\theta^\prime)\cos^2(2\theta)}{8\pi},\\
\mathcal{F}_{5}(\theta,\theta')
&\equiv&\frac{\sin(2\theta^\prime)
\sin(2\theta)}{4\pi},\\
\mathcal{F}_{6}(\theta,\theta')
&\equiv&\frac{\sin(2\theta^\prime)\sin^2(2\theta)}{64\pi},\\
\mathcal{F}_{7}(\theta,\theta')
&\equiv&\frac{\cos(\theta^\prime)
\cos(\theta)}{\pi},\\
\mathcal{F}_{8}(\theta,\theta')
&\equiv&\frac{\cos(\theta^\prime)\cos^2(\theta)}{8\pi},\\
\mathcal{F}_{9}(\theta,\theta')
&\equiv&\frac{\sin(\theta^\prime)
\sin(\theta)}{\pi},\\
\mathcal{F}_{10}(\theta,\theta')
&\equiv&\frac{\sin(\theta^\prime)
\sin^2(\theta)}{8\pi},\label{theta_theta'}
\end{eqnarray}
\end{small}
where $\theta$ and $\theta'$ serve as the momentum directions of
nodal-line fermions involved in the fermion-fermion interactions.

\end{CJK*}

\end{document}